\documentclass[aps,pre,twocolumn,superscriptaddress,showpacs]{revtex4-2}
\usepackage{graphicx}
\usepackage{array}
\usepackage{amsfonts}
\usepackage{amssymb,amsmath,multirow,rotate}
\usepackage{mathrsfs}
\usepackage{booktabs}
\usepackage{threeparttable}
\usepackage{multirow}
\usepackage{epsfig}
\usepackage{threeparttable}
\usepackage{chngpage}
\usepackage{latexsym}
\usepackage{dcolumn}
\usepackage{graphics,graphicx,bm,fleqn,epic,eepic,float}
\usepackage{verbatim}
\usepackage[dvipsnames]{xcolor}
\usepackage{xr}
\usepackage{float}
\usepackage{listings} 
\usepackage{placeins} 
\usepackage{tikz}
\usepackage{amsmath, nccmath}
\usepackage{soul}

\definecolor{red}{rgb}{1,0,0}

\definecolor{blue}{rgb}{0,0,1}

\usepackage{color}
\usepackage[normalem]{ulem}
\definecolor{darkgreen}{rgb}{0.1,.6,.1}

\begin{document}

\title{Anticipating tipping in spatiotemporal systems with machine learning}

\date{\today}

\author{Smita Deb}
\affiliation{School of Electrical, Computer, and Energy Engineering, Arizona State University, Tempe, AZ 85287, USA}

\author{Zheng-Meng Zhai}
\affiliation{School of Electrical, Computer, and Energy Engineering, Arizona State University, Tempe, AZ 85287, USA}

\author{Mulugeta Haile}
\affiliation{DEVCOM Army Research Office, 6340 Rodman Road, Aberdeen Proving Ground, MD 21005-5069, USA}

\author{Ying-Cheng Lai} \email{Ying-Cheng.Lai@asu.edu}
\affiliation{School of Electrical, Computer, and Energy Engineering, Arizona State University, Tempe, AZ 85287, USA}
\affiliation{Department of Physics, Arizona State University, Tempe, Arizona 85287, USA}

\begin{abstract}

In nonlinear dynamical systems, tipping refers to a critical transition from one steady state to another, typically catastrophic, steady state, often resulting from a saddle-node bifurcation. Recently, the machine-learning framework of parameter-adaptable reservoir computing has been applied to predict tipping in systems described by low-dimensional stochastic differential equations. However, anticipating tipping in complex spatiotemporal dynamical systems remains a significant open problem. The ability to forecast not only the occurrence but also the precise timing of such tipping events is crucial for providing the actionable lead time necessary for timely mitigation. By utilizing the mathematical approach of non-negative matrix factorization to generate dimensionally reduced spatiotemporal data as input, we exploit parameter-adaptable reservoir computing to accurately anticipate tipping. We demonstrate that the tipping time can be identified within a narrow prediction window across a variety of spatiotemporal dynamical systems, as well as in CMIP5 (Coupled Model Intercomparison Project 5) climate projections. Furthermore, we show that this reservoir-computing framework, utilizing reduced input data, is robust against common forecasting challenges and significantly alleviates the computational overhead associated with processing full spatiotemporal data.


\end{abstract}


\maketitle

\section{Introduction} \label{sec:intro}

Critical transitions represent one of the most significant challenges in the study of complex nonlinear systems. These are abrupt, often irreversible shifts in a system's state, such as a crisis~\cite{GOY:1982,GOY:1983} at which a chaotic attractor is destroyed and replaced by a chaotic transient leading to a different, often catastrophic state. This phenomenon is not merely a theoretical curiosity; it has real-world implications. In electrical power systems, for instance, voltage collapse can occur after the system enters a state of transient chaos~\cite{DL:1999}. Similarly, in ecology, slow environmental deterioration can push a system into transient chaos, which is then followed by species extinction~\cite{MY:1994,HACFGLMPSZ:2018}. Critical transitions occur across diverse fields, ranging from the collapse of ecosystems and power-grid blackouts to sudden shifts in regional climates and epileptic seizures. The primary difficulty in forecasting these events is that they are typically preceded by subtle, almost imperceptible changes; a system can appear stable and ``normal'' right up until the moment of collapse. Developing reliable methods for detecting early-warning signals and predicting these transitions is therefore a task of paramount importance, offering the potential to mitigate catastrophic outcomes.

The core challenge lies in predicting such a crisis and the subsequent collapse purely from data, especially when the system is still operating in a seemingly normal chaotic regime. In most practical scenarios, the governing equations of the system are unknown, and only measured time series are available. This limitation renders many model-based approaches ineffective. While methods such as sparse optimization can identify system equations from data~\cite{WYLKG:2011,WLG:2016}, they typically require the underlying dynamics to possess a simple mathematical structure with, e.g., a limited number of polynomial or Fourier series terms. Real-world dynamical systems may not meet this condition, making the model-free, data-driven prediction of crises a significant and outstanding problem. To address these challenges, a machine-learning framework known as adaptable reservoir computing~\cite{KFGL:2021a,KFGL:2021b} was recently developed. This approach has been successfully applied to predicting crises in classical chaotic systems, representing an advance over sparse-optimization-based equation-finding methods. For example, it successfully predicts crises for the Ikeda-Hammel-Jones-Moloney optical cavity map~\cite{Ikeda:1979,IDA:1980,HJM:1985} - a system for which sparse optimization methods fail~\cite{KFGL:2021b}. This success underscores its potential as a robust, model-free predictive tool.

While reservoir computing has proven effective for predicting such critical transitions~\cite{KFGL:2021a,KFGL:2021b,PO:2023,KWGHL:2023}, its success typically relies on the system exhibiting oscillatory behavior prior to the transition. The temporal variations in the training time series provide the rich dynamical information necessary for the model to learn the system's underlying rules. However, the phenomenon of tipping, in its original context~\cite{Scheffer:2004,Schefferetal:2009,Scheffer:2010,WH:2010,DG:2010,CLLLA:2012,BH:2012,DVKG:2012}, involves a transition from one stable steady state to another. This is qualitatively different from the more frequently studied critical transitions that arise from an oscillatory state. Predicting a tipping point thus presents a substantially greater challenge. Specifically, prior to tipping, the system resides in a stable steady state characterized by an absence of oscillations in the dynamical variables, thereby depriving the learning algorithm of the necessary dynamic data.

Recent work~\cite{PKMZGHL:2024} proposed a solution for the machine-learning-based prediction of tipping by exploiting dynamic noise. Time series measured from real-world systems are inherently noisy, and these random fluctuations are naturally well-suited for machine-learning training. When developing a predictive framework, synthetic data is often required for validation; in such cases, time series with random perturbations around the deterministic steady state can be generated via stochastic dynamical modeling. While noise may potentially compromise prediction accuracy, it serves a crucial dual purpose here. First, it facilitates a thorough exploration of the phase space by the system's trajectory, which unveils latent dynamical features that would otherwise remain obscured in noise-free conditions. In fact, dynamical and/or measurement noise in the training dataset can be beneficial through a stochastic-resonance mechanism~\cite{ZKL:2023}. Furthermore, the optimal calibration of noise levels can mitigate overfitting and promote generalization, allowing the reservoir computer to adapt to varying data distributions. By utilizing an adaptable reservoir computer designed to accommodate time-varying parameters, it has been demonstrated~\cite{PKMZGHL:2024} that the model can be successfully trained to predict the future occurrence of tipping.

In the aforementioned machine-learning work on predicting tipping~\cite{PKMZGHL:2024}, the underlying system was assumed to be describable by stochastic ordinary differential equations, meaning the spatial context of the system was completely ignored. However, real-world dynamical systems inherently operate across varying spatial scales~\cite{rietkerk2021evasion,lenton2024remotely}. Traditionally, proximity to a tipping point can be inferred from changes in fluctuations and spatial patterns that accompany a loss of stability. For example, the emergence of characteristic spotted vegetation patterns in various ecosystem models indicates an abrupt transition to a homogeneous barren state in response to reduced rainfall~\cite{kefi2014early,pascual2005criticality,majumder2019inferring}. Phenomena with the potential to lead to a tipping transition, such as the decay of the Atlantic Meridional Overturning Circulation~\cite{lenton2008tipping,drijfhout2013spontaneous,TVBFAF:2020,DD:2023,vWKD:2024}, large-scale Amazon dieback~\cite{change2007intergovernmental}, and Arctic sea-ice collapse~\cite{eisenman2009nonlinear}, are inherently driven by spatial feedbacks operating at basin-wide or hemispheric scales. Ecological tipping points, such as lake eutrophication~\cite{wang2012flickering}, coral reef collapse~\cite{nystrom2001spatial}, and forest-to-grassland transitions~\cite{eby2017alternative}, are similarly driven by positive feedbacks in spatial networks. Therefore, incorporating spatial scales is imperative to accurately predict tipping and design effective resilience strategies. This leads to our central question: can adaptable reservoir computing be exploited to anticipate tipping in spatiotemporal dynamical systems?

In this paper, we address this critical question. Given a spatiotemporal dynamical system described by a nonlinear partial differential equation (PDE), a conventional approach is to perform spatial discretization to obtain a set of coupled ordinary differential equations, and then feed the time series from all grid points into a reservoir computer. However, this brute-force approach is often computationally prohibitive or entirely infeasible. Instead, we seek to project the high-dimensional system onto a lower-dimensional manifold that preserves fundamental characteristics, such as the bifurcation type and the timing of onset. For this, we employ nonnegative matrix factorization (NMF)~\cite{gillis2020nonnegative,wang2012nonnegative}, a dimension-reduction method that provides a low-rank, part-based representation of high-dimensional data while preserving interpretable features. By decomposing the system into additive, nonnegative basis components, NMF captures localized structures that reflect the dominant modes of variability in the original system. Unlike other projection-based methods, NMF avoids the mixing of positive and negative contributions, allowing essential patterns, intensities, and coherent structures to be cleanly retained. We test our framework by predicting tipping timing based on spatiotemporal data from various discrete and continuous reaction-diffusion models. We demonstrate that the framework is robust against varying data lengths, data resolutions, and changes in the training data's proximity to the tipping threshold. We also verify that the framework accurately anticipates tipping transitions while correctly identifying no-transition cases, thereby minimizing false positives. Finally, we show that the method can detect tipping points in CMIP5 climate projections with $95\%$ confidence, underscoring its practical effectiveness for forecasting future tipping events.

Notably, when using data augmented with dynamic noise, the training, validation, and hyperparameter optimization are performed exclusively using data from the pre-critical regime. During the testing or prediction phase, the reservoir computer operates as a closed-loop, deterministic dynamical system, forecasting how the system's dynamical climate will evolve in response to the time-varying bifurcation parameter. Because the training process incorporates no data from the post-tipping regime, as such data would be unavailable in a real-world predictive scenario, the reservoir computer cannot predict the detailed system behavior after tipping has occurred. However, the model successfully generates characteristic signals in its output that precede the transition, thereby making the anticipation of tipping possible.

\begin{figure*} [ht!]
\centering
\includegraphics[width=1\linewidth]{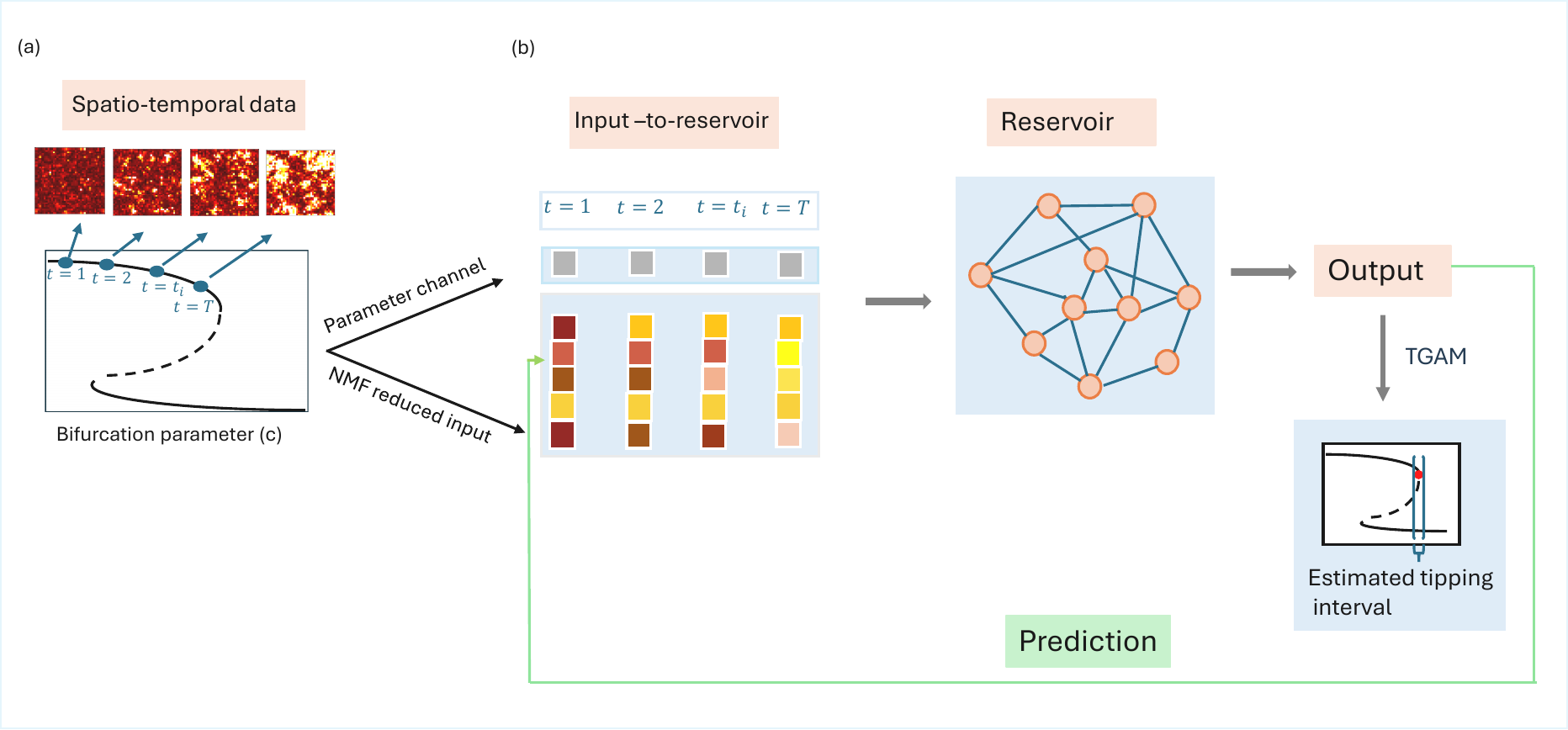}
	\caption{Schematic representation of the parameter-adaptable reservoir-computing framework for estimating tipping points. (a) Spatiotemporal fields (heatmaps) are processed and dimensionally reduced via NMF to form the reservoir input. This is combined with an external parameter channel fed with the bifurcation parameter corresponding to the training samples. (b) Workflow of the parameter-adaptable reservoir computer. The reservoir encodes the input dynamics and produces an output trajectory. A TGAM (Threshold Generalized Additive Model) is then applied to the reservoir output to estimate the timing of the tipping point. Arrows indicate the flow of data from the raw inputs to the final estimated tipping point.}
\label{fig:Schematic}
\end{figure*}

\section{Results}\label{sec:results}

We train and deploy a parameter-adaptable reservoir computer to predict spatiotemporal tipping using synthetic PDE systems and CMIP5 climate projections. Figure~\ref{fig:Schematic} presents a schematic representation of the operational workflow of the framework. During the training and validation phases, the reservoir computer operates in an open-loop configuration, receiving external spatiotemporal data as input. In the testing phase, a closed-loop configuration is employed, where the reservoir computer generates self-sustained predictions by feeding its own output back as input. Additional details regarding the framework and its implementation can be found in the Methods and Supplementary Information (SI).

\subsection{Demonstration of predicting tipping in spatiotemporal systems} \label{subsec:performance}

\begin{figure*} [ht!]
\centering
\includegraphics[width=0.8\linewidth]{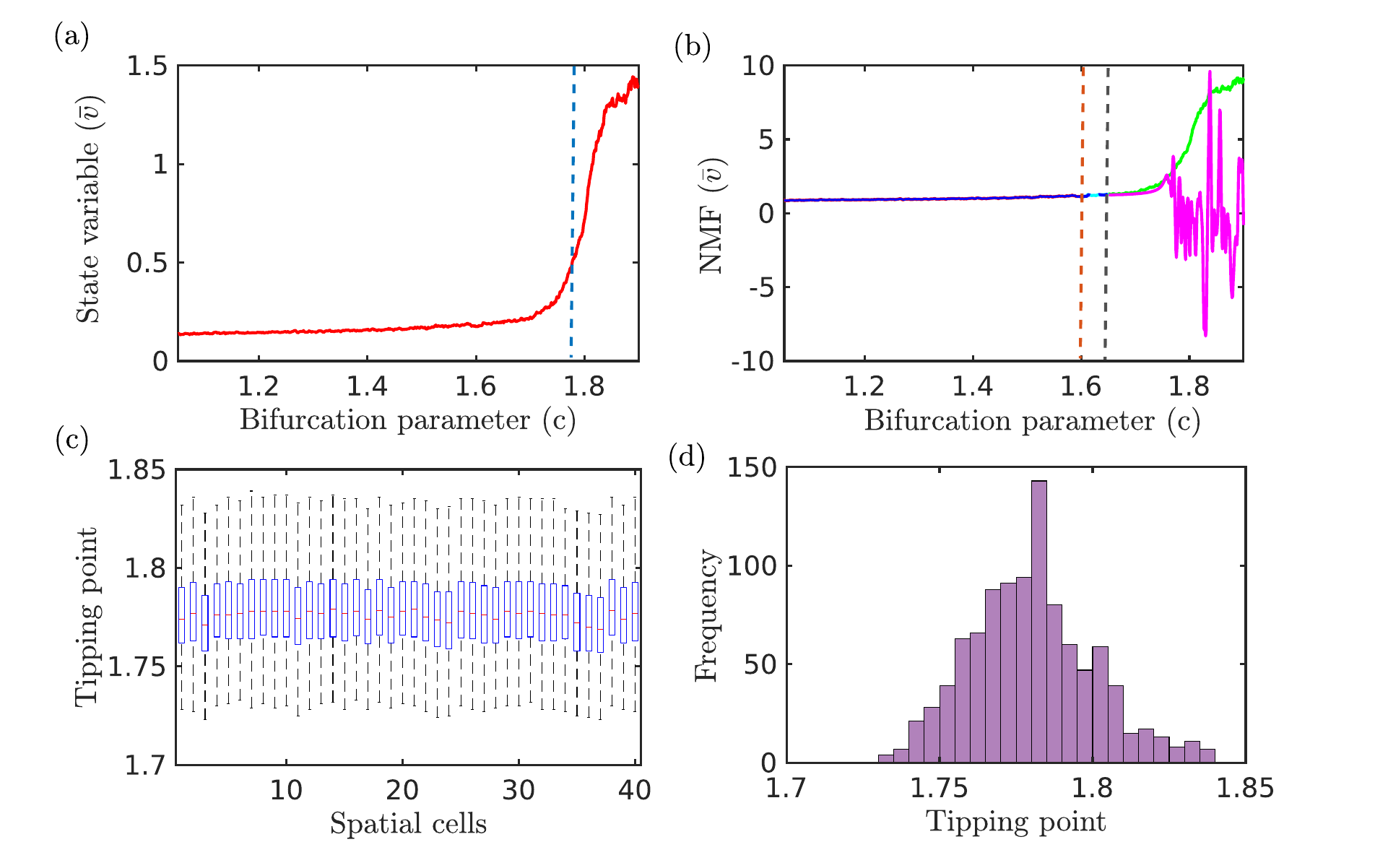}
\caption{Reservoir-computing prediction of the tipping interval in the spatiotemporal vegetation-turbidity model. (a) The spatial mean density of the state variable $v$ versus the bifurcation parameter $c$. The blue dashed curve indicates the tipping time obtained using the TGAM method. (b) Mean-field plots of the training, validation, and testing datasets after applying NMF to the snapshots. The dashed orange and black vertical lines mark the end of training and start of testing, respectively. In the testing phase, the green curve represents the ground truth. (c) Box plots representing the tipping point in each of the $n_{x}(=40)$ spatial cells. (d) A histogram plot displaying the predicted tipping point of the spatiotemporal system from 1000 statistical realizations.}
\label{fig:VT_model}
\end{figure*}

As an illustrative example, we consider spatial data from a one-dimensional model that captures the dynamics of certain ecological systems, known as the vegetation-turbidity model~\cite{carpenter1999management}:
\begin{align} \label{eq:VT}
\dfrac{\partial v}{\partial t}=r+ c\dfrac{v^q}{b+v^q}-sv + D\nabla_{x}^2 v + g(v)\xi_{t}\;,
\end{align}
where the dynamic variable $v$ undergoes a tipping transition when the bifurcation parameter $c$ is gradually varied and reaches a critical threshold value, as indicated by the black dashed line in Fig.~\ref{fig:VT_model}(a). The quantity $\xi$ is a Gaussian white noise process with mean zero and variance $\sigma^{2}$. The bifurcation parameter $c$ is varied linearly as a function of time, making the system non-autonomous. (Further details regarding the model parameters and their physical meanings are provided in Sec.~\ref{subsec:spatiotemporal_model}.) We simulate the spatiotemporal state variable $v$ on an $n_{x} \times n_{y}$ spatial grid at each time step $t$ in the presence of multiplicative Gaussian noise with a small amplitude, generating a time series of snapshots (matrices). Subsequently, we apply NMF to each snapshot to reduce the dimension of the system at each time step $t$ to $n_{x}\times k$ (with $k=1$ in our study). While NMF amplifies the magnitude of the state variable across spatial cells, the underlying system dynamics are preserved. Specifically, the bifurcation structure remains unchanged, and the critical parameter value at which the bifurcation occurs is maintained.

Figure~\ref{fig:VT_model}(a) shows the mean density of the state variable $v$ with respect to the bifurcation parameter $c$. The system exhibits fluctuations within a small neighborhood and remains in the lower stable state until $c$ reaches a threshold value of approximately $1.784$ (see Fig.~\ref{fig:figS3} in SI). At this point, the system undergoes an abrupt transition to an alternate stable state. To predict this behavior, the parameter $c$ is fed into the reservoir through the parameter channel. Figure~\ref{fig:VT_model}(b) displays the mean-field plot for training, validation, and testing on the NMF-reduced spatial data for a single realization. The original input to the reservoir computer consists of 600 snapshots for training, spanning the parameter region $c \in (1,1.6)$, alongside 50 snapshots for validation and testing over 350 steps, which includes the tipping point. (The training, validation, and testing plots for all spatial cells are presented in Fig.~\ref{fig:figS9} in SI.) It can be seen that the reservoir-computing predictions closely follow the original dynamics until the tipping point is reached. Near the transition, the reservoir-generated variable exhibits abnormal behavior, indicating the successful detection of the tipping point. However, because the reservoir computer is not trained on post-tipping data, which possesses a fundamentally different distribution than pre-tipping data, the detailed dynamics beyond the tipping point cannot be anticipated.

Figure~\ref{fig:VT_model}(c) presents box plots of the tipping point for each of the $n_{x}$ spatial cells as predicted by the reservoir computer across $1000$ realizations. The median tipping point among these realizations falls within the interval $(1.75,1.8)$ for the bifurcation parameter $c$. A histogram of the predicted tipping points from the 1000 realizations is shown in Fig.~\ref{fig:VT_model}(d), exhibiting a peak at approximately $1.784$, which agrees with the ground truth. To estimate the tipping time for each instance, we employ the TGAM method~\cite{andersen2009ecological,bestelmeyer2011analysis} (see Sec.~\ref{secS2} in SI for further details).

We quantify the uncertainty in the predicted tipping point using the Wilson confidence interval~\cite{wilson1942confidence}. Specifically, we define a narrow interval $(1.75, 1.8)$ around the true tipping point as the acceptable prediction window. A reservoir prediction is deemed successful if it falls within this interval, and unsuccessful otherwise. Consequently, the collection of reservoir predictions follows a binomial distribution with parameters $(n,p)$, where $n$ denotes the number of test instances and $p$ represents the success probability. The Wilson score interval can be used to quantify the uncertainty in the estimated success rate based on repeated Bernoulli trials, given by:
\begin{align} \label{eq:Wilson_score}
        {\rm CI}_{W}=\frac{p+\frac{z^{2}}{2n}\pm z\sqrt{\frac{p(1-p)}{n}+\frac{z^{2}}{4n^{2}}}}{1+\frac{z^{2}}{n}}, 
\end{align}
where $k$ is the number of predictions falling in the considered tipping interval, $p=\frac{k}{n}$, and $z$ is the standard normal quantile with $z = 1.96$ corresponding to a $95\%$ confidence level. The quantity ${\rm CI}_{W}$ computed from $1000$ realizations of the reservoir computer is approximately $[81,86]$, indicating with $95\%$ confidence that the probability of a prediction lying within the true tipping interval $(1.75,1.8)$ is likely between $81\%$ and $86\%$. Thus, our results confirm that the parameter-adaptable reservoir computer is capable of reliably predicting tipping in spatiotemporal systems.

\begin{figure*} [ht!]
\centering
\includegraphics[width=\linewidth]{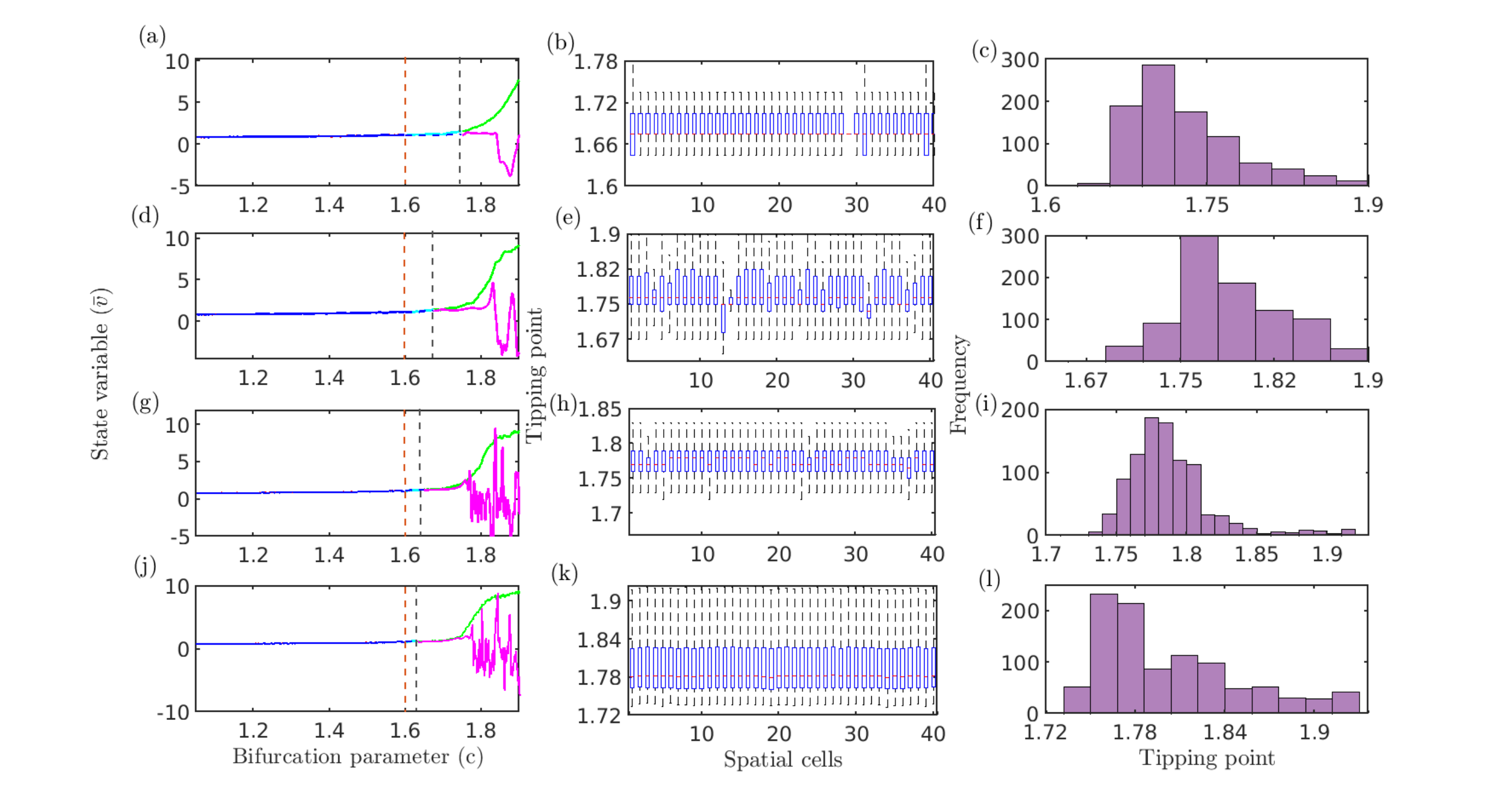}
\caption{Reservoir-computing predictions with varying training data lengths. Panels (a–c), (d–f), (g–i), and (j–l) use 200, 400, 600, and 1000 training snapshots, respectively. Within each triplet, from left to right: mean curves for the training, validation, and testing datasets of the NMF-reduced spatial system; box plots of tipping points for each of the 40 spatial cells; and a histogram of predicted system tipping times from 1000 realizations. Dashed orange and black vertical lines indicate the end of training and the start of testing periods. In the first panel of each row, the green curve represents the ground truth trajectory.}
\label{fig:Performance_VT}
\end{figure*}

\subsection{Robustness against factors influencing detectability of tipping point}

We analyze how variations in the training data length influence the prediction of tipping points. We simulate Eq.~\eqref{eq:VT} using different data lengths of $200$, $400$, $600$, and $1000$ samples over the bifurcation parameter range $c \in (1,1.6)$. Figure~\ref{fig:Performance_VT} shows that performance improves with an increase in data length up to a certain point, beyond which it deteriorates. For a data length of $200$, as shown in Figs.~\ref{fig:Performance_VT}(a-c), the reservoir computer infrequently predicts tipping within the true tipping interval $(1.75,1.8)$, yielding $CI_{W}^{200}=[20,25]$. With longer datasets, the frequency of accurate predictions within the true interval increases, as shown in Figs.~\ref{fig:Performance_VT}(f) and \ref{fig:Performance_VT}(i) with $CI_{W}^{400}=[60,65]$ and $CI_{W}^{600}=[81,86]$, respectively. However, with a further increase in data length---for example, $1000$ samples in the training phase, the confidence interval drops to $CI_{W}^{1000}=[62,68]$, indicating reduced predictive performance.

These results suggest that datasets that are either too short or too long can hinder predictive capabilities. While more samples slow the effective rate of change of the bifurcation parameter and facilitate the learning of the parameter–state relationship, excessive sampling can accumulate noise and redundant information. This leads to an increased risk of overfitting and degraded predictions. Therefore, selecting an optimal training length is critical for accurate tipping-point prediction, much like how other early-warning indicators~\cite{Schefferetal:2009,scheffer2012anticipating,dakos2012methods} require an appropriately chosen sliding window to identify reliable trends.

\begin{figure*}[ht!]
\centering
\includegraphics[width=\linewidth]{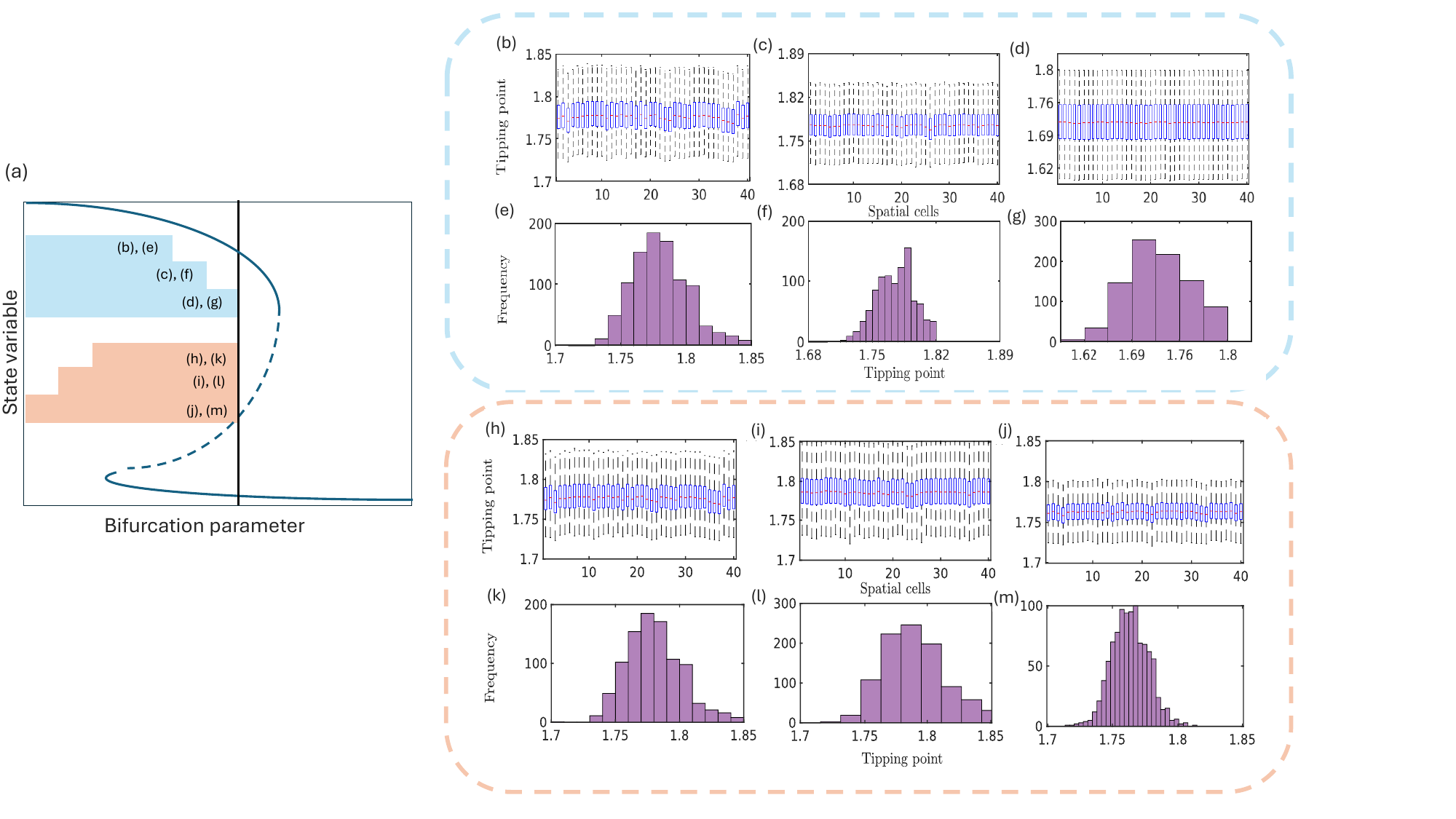}
\caption{Impact of varying the training data on tipping prediction. (a) Choice of training windows by shifting either their end point or their starting point to test the robustness of tipping prediction. Results framed in blue correspond to training windows that terminate at different distances away from the tipping point; those framed in orange correspond to windows whose starting time is progressively moved toward tipping while their length is held fixed. Within each boxed subsection, the top panel displays box plots of the predicted tipping times for all 40 spatial cells, and the bottom panel shows a histogram of system-wide tipping times derived from 1000 reservoir realizations for the setup in (a).}
\label{fig:starting_point}
\end{figure*}

We also investigate how sampling data from different regions of the bifurcation parameter space affects the detectability of tipping points for a fixed number of training samples, as illustrated in Fig.~\ref{fig:starting_point}(a). We sample the training data at varying distances from the tipping point while maintaining a fixed starting point. Figures~\ref{fig:starting_point}(b-g) present the results for training data sampled in the regions $c \in (1,1.6)$, $(1,1.5)$, and $(1,1.4)$, respectively, using a fixed number (600) of training samples. The sample size is fixed at 600, based on the high success rate observed for this data length in the preceding sections. As the distance to the tipping point increases, the predictions deviate further from the true tipping point, as seen in Figs.~\ref{fig:starting_point}(e-g). The peak in the histogram shifts away from the true tipping interval $(1.75,1.8)$, yielding confidence intervals of $CI_{W}=[81,86]$, $[73,78.5]$, and $[52,58]$, respectively. Overall, performance deteriorates as the temporal distance to the tipping point increases.

Conversely, varying the starting point of the training data while keeping the distance to the tipping point fixed has a minimal impact on predictions, as shown in Figs.~\ref{fig:starting_point}(h-m). Specifically, we compare predictions obtained by training the reservoir computer on data sampled from the regions $c \in (1,1.6)$, $(1.1,1.6)$, and $(1.2,1.6)$, respectively, using a fixed set of 600 training samples. In all three cases, the reservoir computer successfully predicts the tipping point with high confidence. The estimated tipping points fall primarily within the true tipping interval, demonstrating that different starting points do not significantly disrupt the model's accuracy. Peaks in the histograms are consistently observed in the true tipping interval, yielding $CI_{W}=[81,86]$, $[78,83]$, and $[71,76.5]$, respectively.

These results align with the theoretical understanding of critical transitions and early warning signals, which dictates that sampling data too far from the tipping point can lead to missing crucial precursor characteristics, resulting in suboptimal predictions. However, our findings also demonstrate that even with limited data availability, the reservoir-computing framework can effectively estimate the tipping point. Overall, the parameter-adaptable reservoir-computing framework is capable of accurately estimating tipping points within a narrow window, even under adverse scenarios involving reduced data length and resolution.

\subsection{Comparison with classical spatial tipping indicators}

\begin{figure*}[ht!]
\centering
\includegraphics[width=0.8\linewidth]{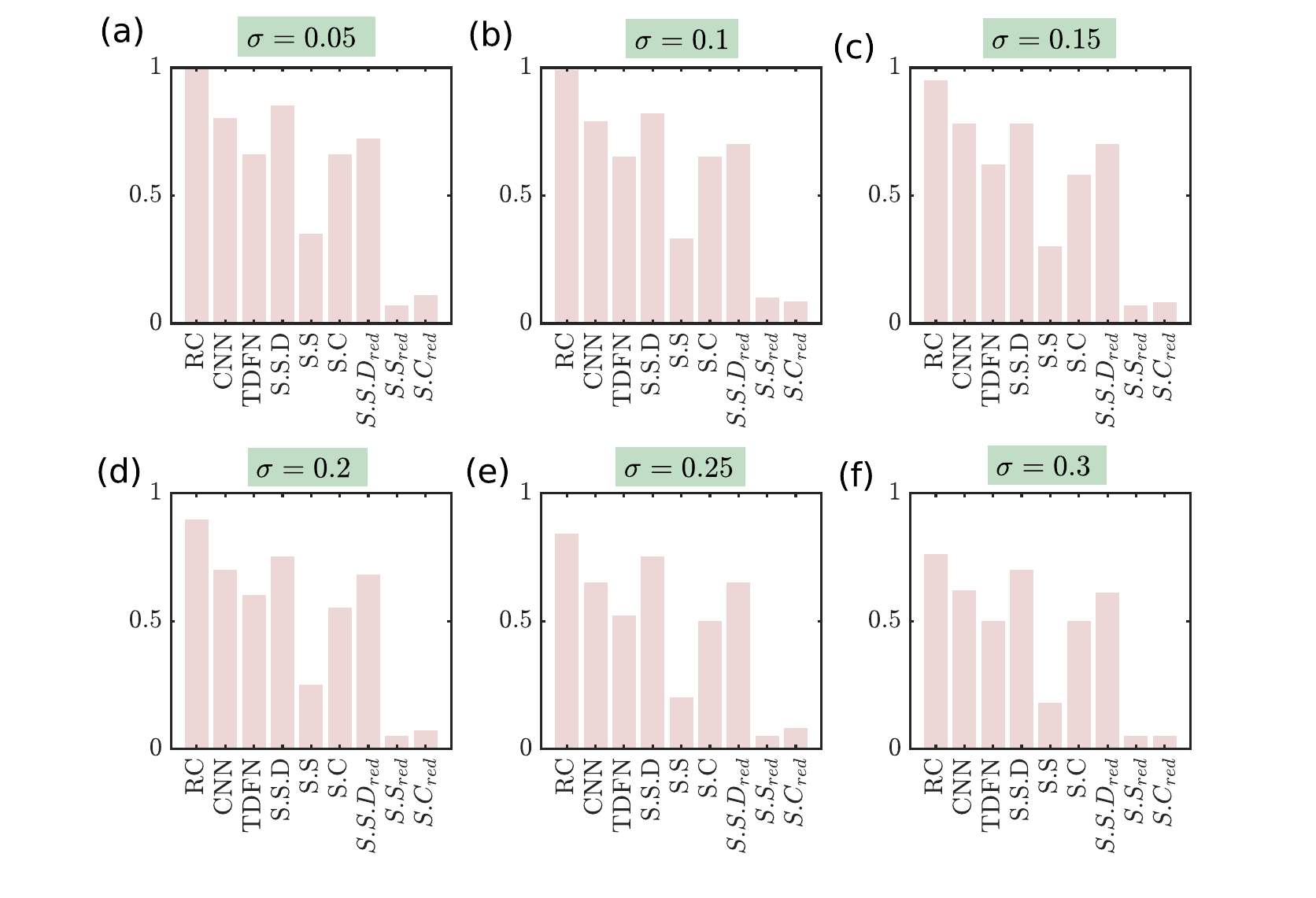}
\caption{Comparison of success rates in tipping point anticipation using reservoir computing, a convolution neural network, a time-delay feedforward network, and classical spatial indicators (spatial standard deviation, spatial skewness, spatial correlation). The spatial data are of size $40\times 40$, obtained from Eq.~\eqref{eq:VT}, with its coarse-grained reduced subsystem subject to multiplicative noise of varying amplitude $\sigma$. Panels (a-f) correspond to $\sigma=0.05, 0.1, 0.15, 0.2, 0.25, 0.3$, respectively.}
\label{fig:figbar}
\end{figure*}

\begin{figure*} [ht!]
\centering
\includegraphics[width=1\linewidth]{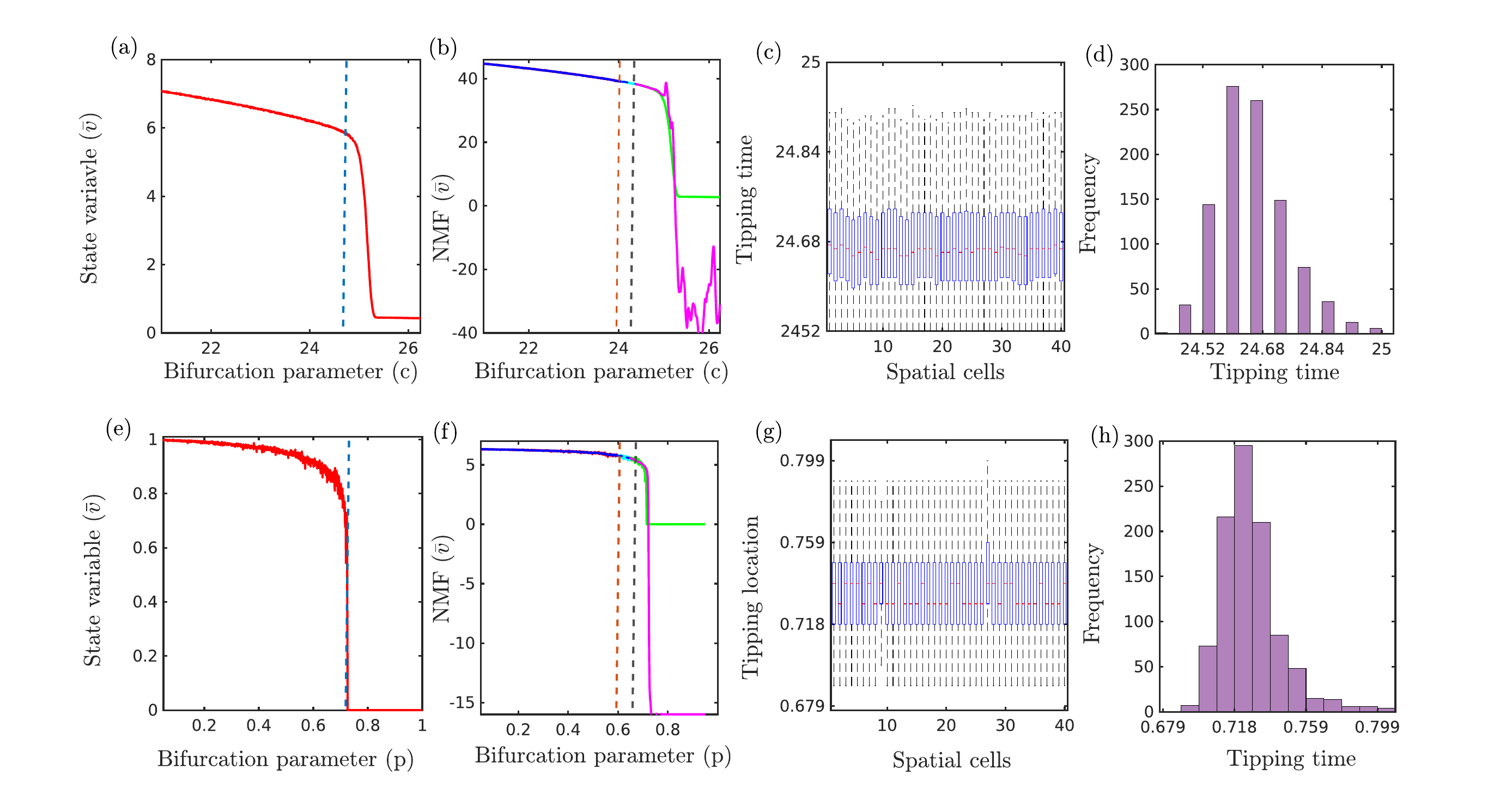}
\caption{Predicting tipping in the spatially extended vegetation grazing system, Eq.~\eqref{eq:VG}, and cellular automata model, Eq.~\eqref{eqCA}. (a) The spatial mean density of the state variable $v$ plotted against a gradually changing bifurcation parameter for the vegetation grazing model. The blue dashed curve marks the ground-truth tipping time identified by the TGAM method. (b) Mean plots for the training, validation, and testing datasets after applying NMF to the snapshots for the vegetation grazing model. The orange and black dashed vertical lines indicate the end of training and the start of testing periods, respectively. In the testing phase, the green curve represents the ground truth trajectory. (c,d) Box plots representing the tipping point for each of the $40$ spatial cells and histogram showing the predicted tipping points based on $1000$ realizations, respectively, for the vegetation grazing system. (e-h) The results corresponding to (a-d) but for the cellular automata model.} 
\label{fig:Grazing}
\end{figure*}

In the search for reliable early warning signals of critical transitions, large fluctuations can mask underlying trends in the data and significantly reduce the effectiveness of standard indicators. We compare the performance of our parameter-adaptable reservoir-computing framework with baseline machine-learning models, namely a convolution neural network (CNN) and a time-delay feedforward neural network (TDFN), as well as with classical spatial indicators used to detect approaching tipping points. The results are summarized in Fig.~\ref{fig:figbar}.

We train the CNN and TDFN models under the same computational setup outlined in Sec.~\ref{subsec:performance} (see SI, Figs.~\ref{fig:fig_fnn}–\ref{fig:fig_cnn}) and find that both models perform poorly in accurately estimating the explicit timing of the tipping point. The closed-loop predictions generated by these models are able to anticipate an impending abrupt transition in a general sense; however, much like classical spatial early warning indicators computed from pre-transition data, they are typically limited to exhibiting qualitative trends that signal an upcoming shift rather than providing a quantitative estimate of the tipping time. In contrast, the reservoir-computing framework proves superior, as it is able to estimate the precise tipping point with high confidence. 

For situations where merely signaling the presence of an upcoming transition is sufficient, we compared the success rates of the machine learning approaches alongside classical spatial early warning indicators (spatial standard deviation, spatial skewness, and spatial correlation), as well as their counterparts computed on the reduced subsystem, across different noise amplitudes. This comparison is vital because real-world data are inherently noisy, and large fluctuations are known to severely degrade the performance of traditional early warning signals.

We consider the system described by Eq.~\eqref{eq:VT} with the same parameterization as in Fig.~\ref{fig:VT_model} and simulate data across different values of the noise amplitude $\sigma$, chosen within a range where tipping occurs. For each noise amplitude, we generate $50$ distinct inputs and evaluate each over $500$ realizations using reservoir computing, TDFN, and CNN, while also computing the classical indicators. The success rate is defined as the fraction of testing instances in which an upcoming transition is correctly anticipated prior to its occurrence, using only pre-transition data. For the machine-learning methods, a realization is counted as a success if the TGAM method estimates a tipping point based on the predicted trajectory, even if that prediction is temporally distant from the actual transition. For the classical spatial indicators, a realization is considered successful if Kendall's $\tau$ statistic exceeds a threshold value ($\tau > 0.5$), indicating a trend stronger than chance.

To ensure a fair comparison, the early warning indicators are computed from the exact same data segment used to train the machine-learning models. We observe that reservoir computing consistently exhibits the best performance among all methods in anticipating tipping across all noise amplitudes, though its success rate naturally decreases as noise increases. This downward trend is observed across all indicators. The CNN and the spatial standard deviation show relatively high success rates and remain robust even at higher noise amplitudes. Notably, the spatial standard deviation computed from the reduced system (obtained by coarse-graining the original system using a $5 \times 5$ submatrix) also performs well. In contrast, the remaining spatial indicators perform poorly when only limited pre-transition data are available. Their performance does improve when data closer to the tipping point is included. To illustrate this behavior, we present a representative plot showing the trends of the spatial early warning indicators in Fig.~\ref{fig:fig_ews}, alongside a table comparing the corresponding Kendall's $\tau$ values, as listed in Tab.~\ref{tab:tab_tau}, computed using both limited data and data extending right up to the tipping point. In the latter case, significantly higher Kendall's $\tau$ values are observed.

We also examined the performance of the reservoir-computing framework in estimating the true tipping point across these different noise levels. A table reporting the Wilson confidence intervals ($CI_W$) at the $95\%$ confidence level for reservoir-computing predictions obtained under varying noise amplitudes is provided in the SI (Tab.~\ref{tab:tab_RC_noise}). Overall, our results demonstrate that adaptable reservoir computing is uniquely capable of estimating the true tipping point over a broad range of noise levels.

\begin{figure*} [ht!]
\centering
\includegraphics[width=1.0\linewidth]{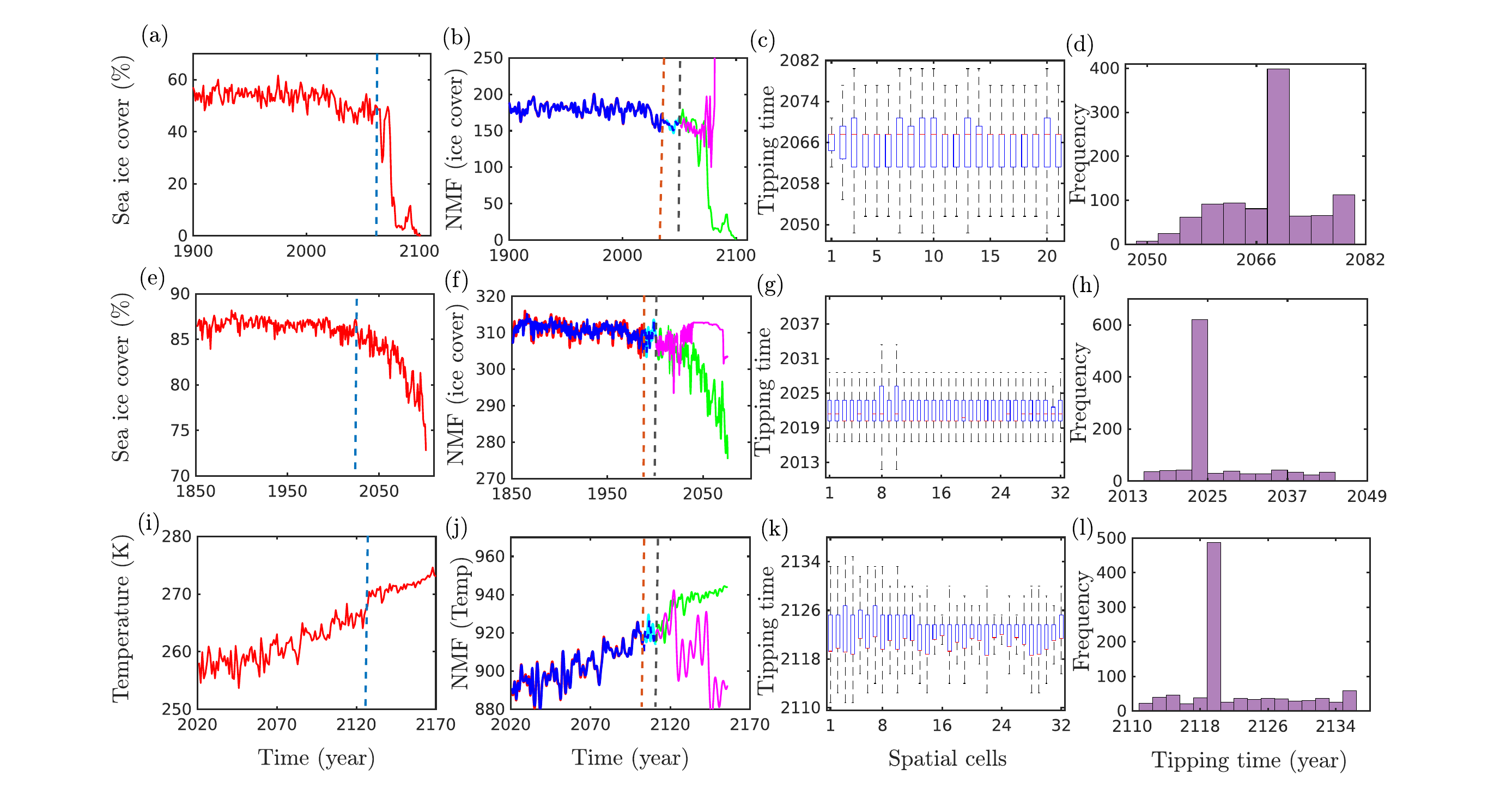}
\caption{Tipping point analysis for CMIP5 climate projections. Data are presented for three variables: (1) annual percentage sea ice cover in the Southern Ocean from MRI‑CGCM3, (2) annual percentage sea ice cover in the Arctic Ocean from CSIRO-MK3-6-0, and (3) air temperature in April from MPI-ESM-LR for the Pacific sector of the Arctic Ocean. (a, e, i) Spatial mean density of the variables (temperature and sea ice cover) over time (in years). The blue dashed curve in each plot marks the ground-truth tipping time as determined by the TGAM method. (b, f, j) Mean plots for the training, validation, and testing datasets of the spatial system after applying NMF to the snapshots. The orange and black dashed vertical lines indicate the end of training and the start of testing periods, respectively. In the testing phase, the green curve represents the ground truth. (c, g, k) Box plots representing the tipping point for each of the $40$ spatial cells. (d, h, l) Histograms showing the predicted tipping points for the three climate systems based on $1000$ realizations.}
\label{fig:CMIP5}
\end{figure*}

\subsection{Tipping prediction in spatial reaction-diffusion models and projected climate data}

In this section, we validate the performance of the parameter-adaptable reservoir computing framework on spatial patterns from two other models: the vegetation grazing (continuous) model~\cite{guttal2009spatial} and the cellular automata (discrete) model~\cite{sankaran2019clustering}, as well as on climate variables from simulations aggregated by the Coupled Model Intercomparison Project 5 (CMIP5). These simulations provide spatiotemporal gridded outputs for a full range of state and flux variables from the major components of the Earth system~\cite{taylor2012overview}. In all these instances, the systems undergo a critical transition, or tipping point, between contrasting states.

\subsubsection{Vegetation grazing system and cellular automata}

Figure~\ref{fig:Grazing}(a) shows the spatial mean abundance of vegetation exhibiting an abrupt shift from a higher to a lower abundance state as the grazing rate increases, which acts as the gradually changing bifurcation parameter. The equation governing the dynamics of the spatially extended vegetation grazing system is given in Eq.~\eqref{eq:VG}. Figures~\ref{fig:Grazing}(b-d) present the framework's predictions and the estimation of the tipping point. The true tipping point of the system, as obtained using the TGAM method, occurs at $c=24.68$ (see Fig.~\ref{fig:figS4} for more details). A distinct peak in the histogram is also observed for the estimated tipping point. Moreover, the computed confidence interval is $CI_{W}=[82,87]$, indicating with $95\%$ confidence that the probability of the estimated tipping point lying within the true tipping interval $(24.65,24.7)$ is between $82\%$ and $87\%$.

Next, we examine the trends in the reservoir-computing predictions using spatial data from a spatially extended ecosystem simulated via a probabilistic update rule on a two-dimensional lattice, also known as the cellular automata model [see Methods, Eq.~\eqref{eqCA}]. Figure~\ref{fig:Grazing}(e) illustrates the spatial mean density as the positive feedback increases in the cellular automata model, which experiences a tipping point when the threshold $p = 0.718$ (see Fig.~\ref{fig:figS5} in SI) is reached. Figures~\ref{fig:Grazing}(f-h) present the predictions and the estimation of the tipping point from $1000$ realizations of the reservoir computer. The histogram exhibits a pronounced peak around $p = 0.72$, which closely matches the actual tipping value of $p = 0.718$. Furthermore, the confidence interval $CI_W = [91, 94]$ indicates that one can be $95\%$ confident that the probability of the estimated tipping point falling within the interval $(0.71, 0.75)$ is between $91\%$ and $94\%$. The training, validation, and testing plots for all spatial cells corresponding to both the vegetation-grazing and cellular automata models are presented in Figs.~\ref{fig:figS10}-\ref{fig:figS11}.

\subsubsection{CMIP5 climate data}

CMIP5 is a global, standardized climate modelling effort coordinated by the World Climate Research Programme~\cite{taylor2012overview}. It provides historical climate simulations ranging from $1850$ to $2005$, which are known to have reproduced observed long-term global climate metrics closely (with correct trends and magnitudes). It also provides future climate projections from $2006$ to $2100$ under different Representative Concentration Pathways, which include the anticipated effects of radiative forcing on the Earth's surface until $2100$. These simulations enable the assessment of how climate variables may evolve under different emissions pathways, supporting scientific evaluation and informing policy decisions regarding climate change mitigation and adaptation. Recent studies have identified tipping behavior in several CMIP5 climate variables, including near-surface air temperature, sea ice, ocean circulation, and land ice across multiple models. From the CMIP5 archive, we selected three test cases with available monthly data under RCP scenarios of future greenhouse gas emissions, each exhibiting clear evidence of tipping in the near future~\cite{drijfhout2013spontaneous,bathiany2020edge}.

We analyze the following three CMIP5 cases: (1) annual percentage sea-ice cover in the Southern Ocean from MRI‑CGCM3 under RCP2.6, (2) annual percentage sea-ice cover in the Arctic Ocean from CSIRO-Mk3-6-0 under RCP8.5, and (3) April near-surface air temperature over the Pacific sector of the Arctic Ocean from MPI-ESM-LR under the RCP8.5 scenario. For each test instance, we apply spatial sampling, extract the target region using a land-ocean mask, and reduce the dimensionality via NMF. This is followed by interpolation across all the $n_x$ cells to generate sufficient training samples. Here, we use 600 training samples for all CMIP5 variables, motivated by our findings from the spatial model data. All three datasets exhibit tipping behavior on an annual timescale, resulting in relatively short effective training series for the reservoir computer.

Because the explicit bifurcation (control) parameter is not directly available in climate projections, and abrupt transitions (tipping points) are observed strictly as a function of time, we make the assumption that the underlying parameter is a linear function of time. We therefore use time as a surrogate parameter. Given a dataset that spans a certain number of years (e.g., from 1900 to 2100), we normalize this actual time interval to the unit interval $[0,1]$ and divide it into training, validation, and testing phases accordingly. We also use interpolation and detrending to generate the sufficient number of samples required for training the reservoir computer, while ensuring that the effective sampling interval remains shorter than the characteristic timescale of the system. Additional details on the CMIP5 variables, the preprocessing steps, and the true tipping time for each variable are provided in Tab.~\ref{tab:tab2} and Figs.~\ref{fig:figS6}-\ref{fig:figS8} in SI.

Figures~\ref{fig:CMIP5}(a, e, i) display the regional mean trajectories of the climate variables as a function of time in years. Following training and validation, we observe that for all three CMIP5 climate variables, the reservoir-computing predictions successfully capture the tipping behavior, with a pronounced, abrupt change occurring at or near the true tipping time, as shown in Figs.~\ref{fig:CMIP5}(b, f, j). The training, validation, and testing plots for all individual spatial cells are presented in Figs.~\ref{fig:figS12}-\ref{fig:figS14} in SI. An abnormal shift is correctly identified in each instance at or close to the tipping point. The box plots in Figs.~\ref{fig:CMIP5}(c, g, k) and the peaks in the histograms in Figs.~\ref{fig:CMIP5}(d, h, l), derived from $1000$ realizations, concentrate tightly around the true tipping time. This indicates strong predictive capabilities of the reservoir computer for anticipating climate tipping points. The corresponding confidence intervals are $CI_{W}=[51,57]$, $[68,74]$, and $[57,63]$ for the April air temperature over the Pacific sector, percentage sea-ice cover in the Arctic, and percentage sea-ice cover in the Southern Ocean, respectively, using a $\pm 2$ year tipping window.

Overall, our results demonstrate that the adaptable reservoir-computing framework reliably predicts spatiotemporal tipping across diverse, complex datasets, remains robust against common forecasting challenges, and imposes minimal computational overhead.

\section{Discussion} \label{sec:discussion}

Anticipating a tipping point and estimating the tipping time in complex nonlinear dynamical systems are difficult tasks due to random fluctuations, uncertainties near bifurcations, and abrupt state changes that occur within narrow parameter ranges or timeframes. Traditional methods for predicting tipping points in temporal and spatial systems typically only provide warnings of impending transitions without estimating a specific tipping window. While recent machine-learning approaches have tackled this issue for dynamical systems and models that do not involve spatial dimensions, incorporating spatial dimensions adds significant complexity and computational demands. Understanding and mitigating tipping points is vital, as timely intervention can prevent critical transitions in systems ranging from the Earth's climate to human organ failure, which are best represented in both space and time. Here, we have demonstrated that adaptable reservoir computing can be exploited to predict tipping points in spatiotemporal dynamical systems, thereby addressing a significant and longstanding challenge.

Parameter-adaptable reservoir computing is effectively an augmentation of conventional reservoir computing, utilizing a parameter channel through which the bifurcation parameter is passed concurrently with the input. As a result, the reservoir computer functions essentially as a parameterized digital twin of the target dynamical system. This makes it possible to forecast system behavior for different parameter values by capturing the parameter–state relationship and identifying the critical parameter threshold, or the tipping point (see Fig.~\ref{fig:fig_nopar}). Using this framework, coupled with the dimension reduction of spatial data via NMF, we have succeeded in estimating tipping points across a number of benchmark spatiotemporal systems.

There are a few issues warranting further discussion. The first concerns the effectiveness of the NMF method. For reducing the dimension of spatiotemporal dynamical systems, linear methods such as principal component analysis (PCA) and independent component analysis, as well as nonlinear methods such as kernel PCA and autoencoders~\cite{ghojogh2023elements}, are commonly used. Some of these methods successfully preserve the original dynamics. The essence of NMF is to factorize the matrix into non-negative components through an iterative update rule followed by optimization, offering high interpretability and yielding representations that remain closely tethered to the original dynamics. The only assumption required is that the matrix entries corresponding to system snapshots at each time instance are non-negative, a condition that is naturally satisfied by the data from our benchmark systems. While alternative methods may also be employed, their suitability depends heavily on the specific preprocessing requirements of the dataset.

The second issue involves a ``null'' test: if the target system exhibits no tipping, will the parameter-adaptable reservoir computer correctly predict that there is indeed no tipping? This is important for ensuring that the machine-learning framework is a reliable estimator. To test this, we generated data that exhibit fluctuations about the mean but do not undergo a critical transition as the bifurcation parameter is varied. The prediction results correctly indicate no such transition, as detailed in Sec.~\ref{secS1} of the SI (Figs.~\ref{fig:figS1}-\ref{fig:figS2}), demonstrating that the reservoir computer is capable of distinguishing between situations where tipping occurs and those where it does not.

The third issue concerns the ``supervised'' nature of parameter-adaptable reservoir computing, where associating the data with specific values of the bifurcation parameter is an essential requirement~\cite{KFGL:2021a,KFGL:2021b,PKMZGHL:2024}. In real-world applications, the underlying bifurcation parameter is often unknown or inaccessible. In such cases, it is not possible to strictly associate the collected training data with a specific parameter value; the CMIP5 projected climate data studied in this paper belongs to this category. Our solution is to adopt the reasonable assumption that the underlying bifurcation parameter varies linearly with time. Thus, a ``time'' variable, appropriately normalized based on the time span of the available training and testing data, can serve as a substitute for the parameter. For the CMIP5 data, we tested three projected climate variables. In each case, the reservoir computer correctly predicts the tipping point within a $\pm 2$ year tipping window. This solution may be of particular importance to climate applications, as well as other fields where available remote sensing data are recorded on a geospatial lattice without any knowledge of the true bifurcation parameter. While such data are often limited in quantity and subject to large random fluctuations, interpolation and preprocessing can be used to alleviate these difficulties to some extent. We also note that an unsupervised machine-learning scheme was recently developed to anticipate critical transitions without explicit knowledge of the bifurcation parameter~\cite{PKGHL:2026}, utilizing a variational autoencoder to extract the parameter directly from the data. It was demonstrated that when this extracted parameter is fed into a parameter-adaptable reservoir computer, crisis bifurcations in different benchmark chaotic systems can be accurately predicted. It remains to be seen if this unsupervised learning strategy can be successfully exploited to anticipate spatiotemporal tipping.

We have also analyzed the robustness of the reservoir-computing framework with respect to limited data and the ``distance'' of such data from the tipping point in the parameter space, both of which pose significant challenges to existing early warning indicators. For example, when the available data are limited, we observe that the reservoir computer can still estimate tipping, albeit less accurately, when trained on as few as $200$ data samples. Our tests point to the existence of an optimal data length for accurate tipping point estimation, a phenomenon that remains to be fully explored and understood. Furthermore, the detectability of a tipping point depends heavily on the parameter regions from which the data are sampled. As we move the training window further away from the tipping point, the reservoir computer can still predict tipping, but with deteriorating performance characterized by a narrowing of the confidence interval. To improve predictions, additional latent features that arise closer to the tipping point must be exploited. Importantly, this is not a limitation of reservoir computing itself, but rather a generic characteristic of dynamical data undergoing a tipping process.

Finally, in scenarios where merely anticipating the occurrence of a tipping point is sufficient, we find that machine-learning-based approaches, even when applied to dimension-reduced systems, perform comparably to, or better than, traditional spatial statistical indicators. For example, while spatial standard deviation consistently exhibits an increasing trend across all scenarios, this increase alone is not sufficient to reliably signal an impending tipping point. Such growth in standard deviation may also arise from strong stochastic forcing, external perturbations, or transitions that do not ultimately involve a tipping event. In contrast, machine-learning methods provide more robust and interpretable predictions by implicitly learning system-specific dynamical features. Among these, parameter-adaptable reservoir computing consistently outperforms other approaches by successfully capturing the complex interplay between the parameter and the system state. Overall, our results demonstrate that, when coupled with dimension-reduced spatial data, reservoir computing offers an effective and generalizable framework for estimating tipping points in complex spatiotemporal dynamical systems.

\section{Methods} \label{sec:Methods}

\subsection{Spatiotemporal models exhibiting tipping} \label{subsec:spatiotemporal_model}

A common class of spatially extended systems that undergo critical transitions are those modeled by stochastic reaction–diffusion partial differential equations driven by spatiotemporal white noise:
\begin{align} \label{eq:DiffuE}
\dfrac{\partial v}{\partial t}=f(v,c)+D{\nabla}^{2}_{\mathbf{x}}v+ g(v)\xi_{t},
\end{align}
where $v \equiv v(\mathbf{x},t)$ is the scalar field at the two-dimensional spatial coordinate $\mathbf{x}$ and time $t$, $f(v,c)$ denotes the deterministic nonlinear function governing the system's local dynamics, $c$ is a bifurcation or driving parameter, $\nabla^2_{\mathbf{x}}$ is the Laplacian operator, and $D$ is the diffusion constant. The last term in Eq.~(\ref{eq:DiffuE}) models stochastic forcing or noise, where $g(v)$ is the coupling function between the field and the noise, and $\xi_{t}$ is a random process with zero mean and variance $\sigma^{2}$ (where $\sigma$ measures the noise amplitude). The noise is considered multiplicative if the function $g(v)$ is not constant, and additive if $g(v)$ is constant. If the deterministic dynamics are bistable, tipping can occur through a saddle-node bifurcation.

The first spatiotemporal system used to test our reservoir-computing predictions is the vegetation-turbidity model, given by Eq.~(\ref{eq:VT}). Originating in marine ecology, it describes the dynamics of lake eutrophication caused by excess nutrients~\cite{carpenter1999management}. Here, $v$ is the nutrient concentration in the water column, $r$ is the water input rate, $s$ is the rate of loss of $v$, $c$ denotes the recycling rate of $v$ by consumers or other factors such as sedimentation, and $g(v)=v$ corresponds to the density-dependent factor that modulates the white noise $\xi_{t}$. The nonlinear term accounts for the total recycling of $v$ and is assumed to follow a sigmoidal curve where admissible values of $q$ are $q\geq 2$ (we set $q = 2$ in our numerical simulations), and $b$ is the half-saturation constant. The specific parameter values are $r = 0.1$, $b = 1$, $s = 1$, $\sigma=0.15$, and $c \in [1, 2]$. The initial density in each spatial cell is sampled randomly from the interval $[0.4,0.5]$, and the system is solved numerically using a standard forward-time, centered-space scheme. A critical transition, or tipping point, characterized by an abrupt shift from one stable state to a contrasting state following a small change in the bifurcation parameter across a threshold, is a hallmark phenomenon of lake eutrophication~\cite{scheffer2003catastrophic}. The model in Eq.~\eqref{eq:VT} also describes the autoactivating switching of gene expression~\cite{weber2013stochastic} between alternate stable states, which can model the onset of irreversible conditions such as cancer, epileptic seizures, and diabetes.

The second system we study is a continuous population model in the presence of grazing pressure~\cite{guttal2009spatial}. This model has been used to study a variety of systems, such as vegetation in semiarid ecosystems~\cite{noy1975stability}, the exploitation of fish communities~\cite{steele1984modeling}, and spruce budworm dynamics~\cite{may1977thresholds}. In our study, a spatially extended version of this model, known as the vegetation-grazing model, is used. It describes the transition from a vegetated state to an overexploited state as grazing pressure crosses a critical threshold. The model is expressed as:
\begin{align} \label{eq:VG}
\frac{\partial v}{\partial t}=rv\bigg(1-\frac{v}{v_c}\bigg)-\frac{c v^2}{v^2+1}+D\nabla^2_{x}v+\xi_t \frac{v^2}{v^2+1},
\end{align}
where $v$ is the vegetation density and $\xi_t$ represents multiplicative white noise. The other parameters are: maximum growth rate $r=10$, carrying capacity $v_c=10$, and maximum grazing rate $c \in [20, 27]$. The initial population $v_{in}$ across all cells is set to the same state and is sampled from the interval $[10,10.1]$.

For both models, described by Eqs.~(\ref{eq:DiffuE}) and (\ref{eq:VG}), the diffusion/dispersion rate is set to $D=0.001$, the spatial resolution is $dx=dy=0.1$, the spatial grid size is $n_x\times n_y = 40 \times 40$, and the time increment is $dt=10^{-3}$.

We also analyze a discrete spatiotemporal system, known as the cellular automata model, to validate the general applicability of our reservoir-computing prediction framework. Specifically, the dynamics of the model are described by the following set of local birth-death processes~\cite{sankaran2019clustering}:
\begin{align} \nonumber
01 \rightarrow 11 & =10 \rightarrow 11=p, \\ \nonumber
01 \rightarrow 00 & =10 \rightarrow 00=1-p, \\ \label{eqCA}
11 \rightarrow 01 & =11 \rightarrow 10=(1-p)(1-q), \\ \nonumber
110 \rightarrow 111 & =011 \rightarrow 111=q,
\end{align}
where cells possess binary states, with $0$ denoting an unoccupied state and $1$ denoting an occupied state. At each time step, randomly selected cells are updated according to a local establishment probability $p$, which serves as the bifurcation parameter, modulated by local positive feedback via the parameter $q$. The parameter $q$ enhances birth rates when neighboring cells are occupied and increases death rates when neighbors are unoccupied. Varying $q$ can lead to a critical transition that occurs for $q \geq 0.85$. In our simulations, we set $q=0.92$ and use a two-dimensional lattice of $40 \times 40$.

For each model, we generate sequences of spatiotemporal snapshots of dimension $n_{x}\times n_{y}$ for training, validation, and testing the reservoir computer. However, processing high-dimensional spatiotemporal data typically requires a single, prohibitively large reservoir or a massive ensemble of small reservoirs operating in parallel~\cite{PHGLO:2018}. Dimension reduction that preserves the essential spatiotemporal dynamics, and ensures the critical bifurcation threshold is retained without distortion, can significantly reduce computational complexity.

Generally, dimension-reduction methods~\cite{ghojogh2023elements} seek a compact representation of high-dimensional matrix data that preserves its essential structure, variance, or interpretability. Among these, linear methods such as principal component analysis (PCA) or singular-value decomposition project data onto orthogonal directions to capture maximal variance. Nonlinear approaches, such as kernel PCA, diffusion maps, and isomaps, recover low-dimensional geometry when the data lie on a curved manifold. Matrix factorization methods decompose data into interpretable components under constraints such as nonnegativity or sparsity. While linear methods often fail to capture the complex patterns inherent in spatial data exhibiting tipping, nonlinear maps can be difficult to implement. Matrix-factorization methods, however, offer an efficient framework for reducing high-dimensional data into a low-rank representation, retaining dominant patterns in the spatial data with high interpretability. In our study, we use NMF~\cite{gillis2020nonnegative,wang2012nonnegative}, which operates by factorizing any non-negative matrix $\mathbb{X}$ of dimension $m\times n$ into non-negative matrices $\mathbb{W}, \mathbb{H} \geq 0$ such that $\mathbb{X} \cong \mathbb{W}\mathbb{H}$. Here, $\mathbb{W} \in \mathcal{R}^{m\times k}$ is the basis matrix and $\mathbb{H} \in \mathcal{R}^{k \times n}$ is the coefficient matrix. The matrices $\mathbb{W}$ and $\mathbb{H}$ are obtained by applying iterative update rules:
\begin{subequations}
\begin{align}
H_{ij} \leftarrow H_{ij}\frac{(\mathbb{W}^{T}\mathbb{X})_{ij}}{(\mathbb{W}^{T}\mathbb{W}\mathbb{H})_{ij}}\\
W_{ij} \leftarrow W_{ij}\frac{(\mathbb{X}\mathbb{H}^{T})_{ij}}{(\mathbb{W}\mathbb{H}\mathbb{H}^{T})_{ij}}\
\end{align}
\end{subequations}
followed by optimizing ${\rm min}_{\mathbb{W}\mathbb{H}}||\mathbb{X}-\mathbb{W}\mathbb{H}||^{2}$. In our work, we set $m = n_x$ and $k=1$, such that the input to the reservoir computer at each time step is a reduced vector of dimension $n_x$.

\subsection{Parameter-adaptable reservoir computing} \label{secM3}

We employ parameter-adaptable reservoir computing~\cite{KFGL:2021a,PKMZGHL:2024} to estimate spatiotemporal tipping points. A reservoir computer typically consists of three layers: an input layer, a hidden layer, and an output layer. The input layer maps a low-dimensional input signal into a high-dimensional hidden state $\mathbf{r}(t)$ via an input weight matrix $\mathbb{W}_{\rm in}$, and the output layer maps $\mathbf{r}(t)$ back to the desired target space via the output matrix $\mathbb{W}_{\rm out}$. The reservoir (hidden layer) contains $N$ mutually coupled, one-dimensional dynamical units, or nodes. Stacking the state of each node at time $t$ forms the hidden state vector $\mathbf{r}(t)$. The reservoir nodes constitute a complex network whose connection topology is characterized by an adjacency matrix $\mathbb{A}$ with randomly and independently chosen elements. The link density $d$ of the network is a hyperparameter fixed during training. The connection matrix $\mathbb{A}$ is rescaled so that the resulting matrix has negative eigenvalues and a prescribed spectral radius $\rho$, the magnitude of the largest eigenvalue of $\mathbb{A}$, which plays a key role in maintaining the overall stability and memory capacity of the reservoir.

A parameter-adaptable reservoir computer augments the conventional reservoir structure with an auxiliary input parameter channel, which is connected to all reservoir nodes through a matrix $\mathbb{W}_{c}$. The elements of the input weight matrix $\mathbb{W}_{\text{in}} \in \mathcal{R}^{N \times n_{x}}$ and the parameter weight matrix $\mathbb{W}_{\text{c}} \in \mathcal{R}^{N \times 1}$ are sampled uniformly from the range $[-\gamma,\gamma]$, where $n_{x}$ is the dimension of the input vector. The readout matrix $\mathbb{W}_{\rm out}$ then maps $\mathbf{r}(t)$ to the output layer, yielding the predicted output vector $\mathbf{z}(t)$. The dynamics of the parameter-adaptable reservoir computer are governed by:
\begin{align} \label{eq:RV}
\mathbf{r}(t+\Delta t) &=(1-\alpha)\mathbf{r}(t) \\ \nonumber
	& +\alpha \tanh \big(\mathbb{A}\cdot\mathbf{r}(t)+\mathbb{W}_{in}\cdot\mathbf{v}(t)+k_{c}\mathbb{W}_{c}\cdot (c+b_{0}) \big), \\ \label{eq4}
	\mathbf{z}(t) &=\mathbb{W}_{out} \cdot \mathbf{f}\big(\mathbf{r}(t)\big),
\end{align}
where $\alpha$ is the leakage parameter, $\Delta t$ is the time step, $\tanh(\cdot)$ is the hyperbolic tangent function serving as the nonlinear activation function for the artificial neurons, $c$ is the bifurcation parameter, $k_{c}$ is the scaling factor, $b_{0}$ is the bias, and $\mathbf{v}$ is the NMF-reduced spatial input. The nonlinear output function takes the form:
\begin{align}
	f_i\big(\mathbf{r}(t)\big) = \begin{cases}
r_{i}^2, & i \ \text{odd},\\[6pt]
r_{i},   & i \ \text{even}.
\end{cases}
\end{align}
where $i$ is the index of the reservoir nodes. The elements of the matrices $\mathbb{W}_{\rm in}$, $\mathbb{W}_{\rm c}$, and the reservoir connectivity matrix $\mathbb{A}$ are generated randomly and remain fixed during training. However, $\mathbb{W}_{\rm out}$ is optimized through training such that, given past dynamical variables as input, the reservoir can accurately forecast the future evolution of the target system. This straightforward training process endows the reservoir computer with high computational efficiency. The evolution of the reservoir states follows Eq.~\eqref{eq4}.

The parameter-adaptable reservoir computer relies on three foundational principles: (i) the fading memory property, whereby the reservoir state at time $t$ depends primarily on recent inputs while the influence of distant inputs decays; (ii) expressive embedding, mapping low-dimensional inputs into a rich, high-dimensional state space so a linear readout can easily extract task-relevant features; and (iii) bounded, stable reservoir dynamics. A reservoir computer satisfying these properties has proven effective for the model-free, data-driven prediction of chaotic time series.

During training, the NMF-reduced input $\mathbf{v}$ is passed concurrently with the bifurcation parameter $c$ into the reservoir computer. The optimal output matrix $\mathbb{W}_{\rm out}$ can be conveniently obtained through ridge regression:
\begin{align}
\mathbb{W}_{\text{out}} = \mathbb{Y}\mathbb{R}^T (\mathbb{R}\mathbb{R}^T + \beta \mathbb{I})^{-1} ,
\end{align}
where $\mathbb{Y}$ is the matrix of target output vectors, and $\beta$ is a hyperparameter controlling the amount of regularization. All hyperparameters are determined individually for each target system and the CMIP5 climate projected data through Bayesian optimization, with the selected values presented in Tab.~\ref{tab:tab1} (see SI Sec.~\ref{secS3}).

During the training and validation phases, the reservoir operates in an open-loop configuration, where the true signal continuously drives the network. This ensures that the internal dynamics can synchronize with those of the target system, facilitating an accurate estimation of the readout weights. For prediction, the system is switched to a closed-loop (autonomous) mode, in which the reservoir is driven purely by its own output and evolves without external input. This self-sustained feedback mechanism enables long-term autonomous forecasting. Ultimately, the parameter-adaptable reservoir computer demonstrates the ability to reliably and efficiently predict tipping events in complex spatiotemporal systems.

\section*{Acknowledgements}

This material is based upon work supported by the Laboratory University Collaboration Initiative (LUCI) program, through an award made by the Office of the Under Secretary of War for Research and Engineering \big(OUSW(R\&E)\big), Science and Technology (S\&T)/Foundations. This work was also supported by the US Army Research Office under Grants No.~W911NF-24-2-0228 and No.~W911NF-26-2-A002.

\section*{Author contributions}

S.D., Z.-M.Z., M.H. and Y.-C.L. designed the research project, the models, and methods. S.D. performed the computations. S.D., Z.-M.Z., M.H. and Y.-C.L. analyzed the data. S.D., Z.-M.Z., and Y.-C.L. wrote the paper. Y.-C.L. edited the manuscript.

\section*{Declarations}

The authors declare no competing interests.

\section*{Data availability} 

Codes and data are available in a Github repository (\url{https://github.com/SMITA1996/Adaptable-rc_spatiotemporal}).

\section*{Additional information}

Supplementary information is enclosed below.

\section*{Correspondence}

Correspondence and requests for materials should be addressed to Ying-Cheng.Lai@asu.edu.

\bibliography{STipping}

\newpage

\appendix

\begin{widetext}

\renewcommand{\thesection}{S\arabic{section}}  
\renewcommand{\thetable}{S\arabic{table}}  
\setcounter{table}{0}
\renewcommand{\thefigure}{S\arabic{figure}}
\setcounter{figure}{0}
\renewcommand{\theequation}{S.\arabic{equation}}
\setcounter{equation}{0}

\section*{Supplementary information}

\begin{figure*} [ht!]
\centering
\includegraphics[width=\linewidth]{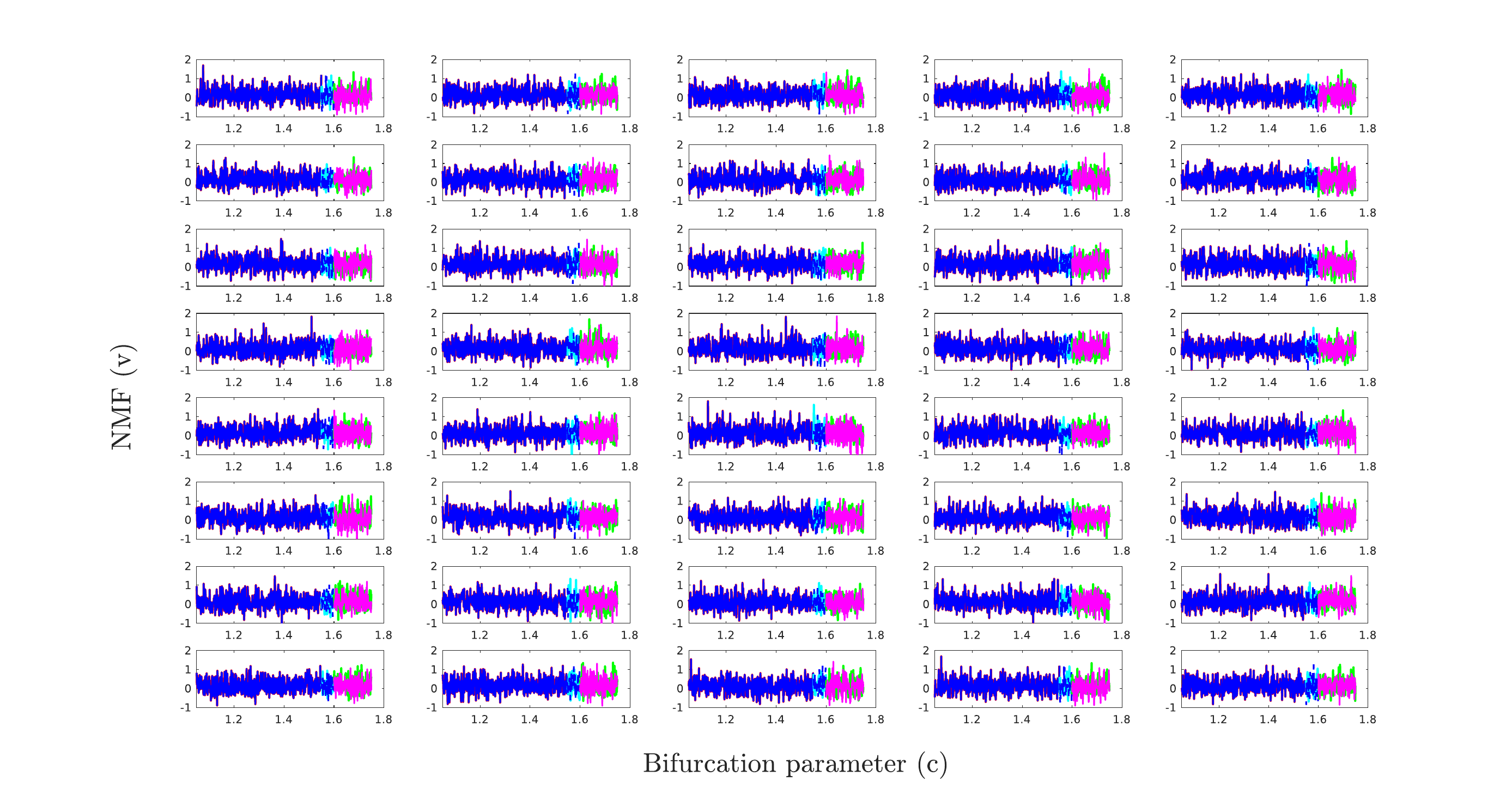}
\caption{Test of false positives. Shown is the NMF-reduced density of the state variable $\mathbf{v}$ versus a gradual change in the parameter $c$ for each spatial cell. Plots are displayed for the training, validation, and testing datasets of the spatial system following the application of NMF to the data snapshots. The training dataset covers the parameter range $c \in (1,1.55)$, validation for $c \in (1.55,1.6)$, and testing is performed on the remaining parameter regime. The reservoir computer predicts no transition in each individual cell, attesting to its ability to eliminate false positives.}
\label{fig:figS1}
\end{figure*}

\begin{figure} [ht!]
\centering
\includegraphics[width=0.6\linewidth]{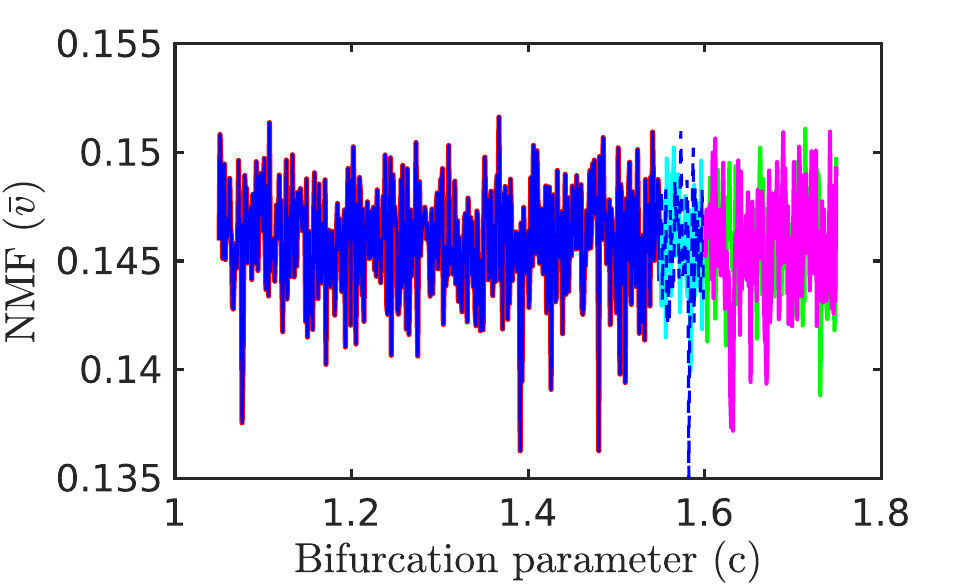}
\caption{The spatial mean density of the state variable $\mathbf{v}$ corresponding to the plots in Fig.~\ref{fig:figS1}. The reservoir computer robustly produces true negatives.}
\label{fig:figS2}
\end{figure}

\section{Test of false positive} \label{secS1}

In the main text, it was demonstrated that a properly trained parameter-adaptable reservoir computer can accurately estimate tipping points. Confidence in this result can be further established by assessing false positives. That is, if a target system does not exhibit any tipping behavior, would the reservoir computer output a false prediction of a tipping event? To test this, we employed data generated from the vegetation–turbidity system, Eq.~\eqref{eq:VT}, but in a different parameter regime that exhibits sustained fluctuations around a mean without any tipping or abrupt transition. As shown in Figs.~\ref{fig:figS1}-\ref{fig:figS2}, the reservoir computer correctly predicts the absence of tipping, demonstrating its robustness against false positives.

\section{Importance of parameter channel}

Figure~\ref{fig:fig_nopar} shows the prediction results for $\mathbb{W}_{c}=0$ in Eq.~\eqref{eq:RV}, a scenario where the parameter channel is effectively nullified. It can be seen that the reservoir-computing predictions do not signal any abrupt transition, attesting to the critical importance of the parameter channel in successfully anticipating tipping points.

\begin{figure} [ht!]
\centering
\includegraphics[width=0.5\linewidth]{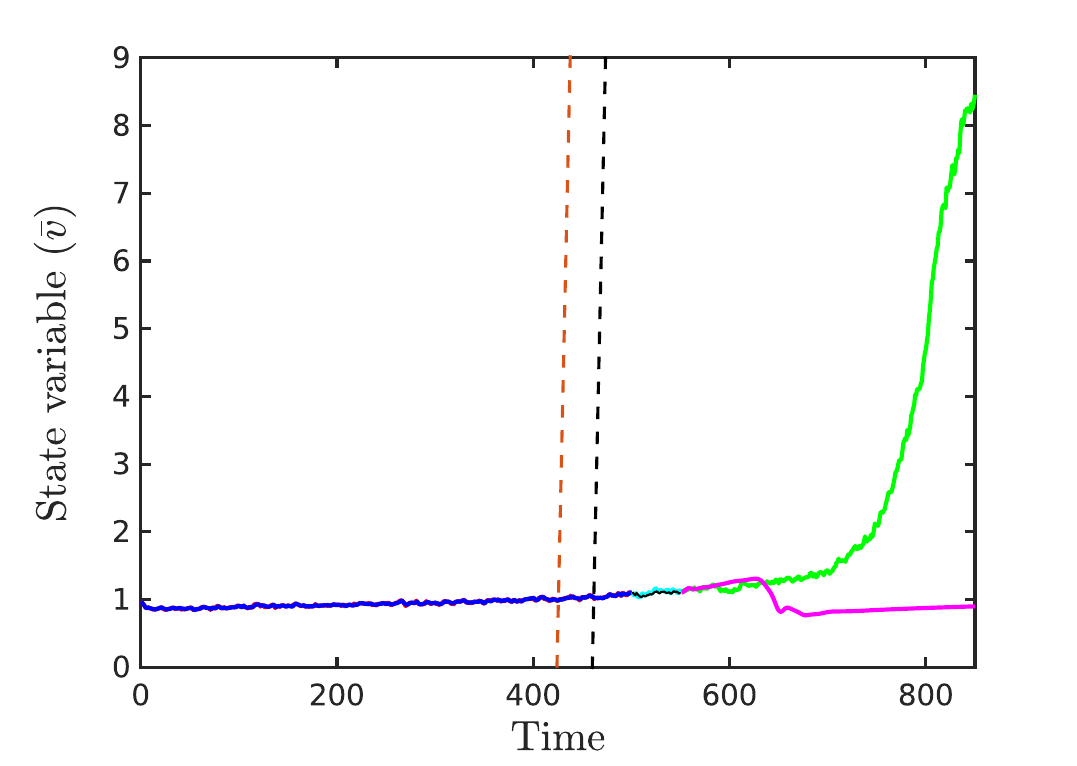}
\caption{Prediction of tipping in the vegetation-turbidity model, Eq.~\eqref{eq:VT}, when the parameter channel of the reservoir computer is nullified. Shown are the mean fields for the training, validation, and testing datasets of the system after applying NMF to the snapshots. The dashed orange and black vertical lines mark the end of the training and the start of the testing periods, respectively. In the testing phase, the green curve represents the ground truth. The setup is identical to Fig.~\ref{fig:VT_model}, except that it does not involve a parameter channel receiving the bifurcation parameter values corresponding to the input during training.}
\label{fig:fig_nopar}
\end{figure}

\section{Baseline methods for anticipating tipping}

\begin{figure} [ht!]
\centering
\includegraphics[width=1.0\linewidth]{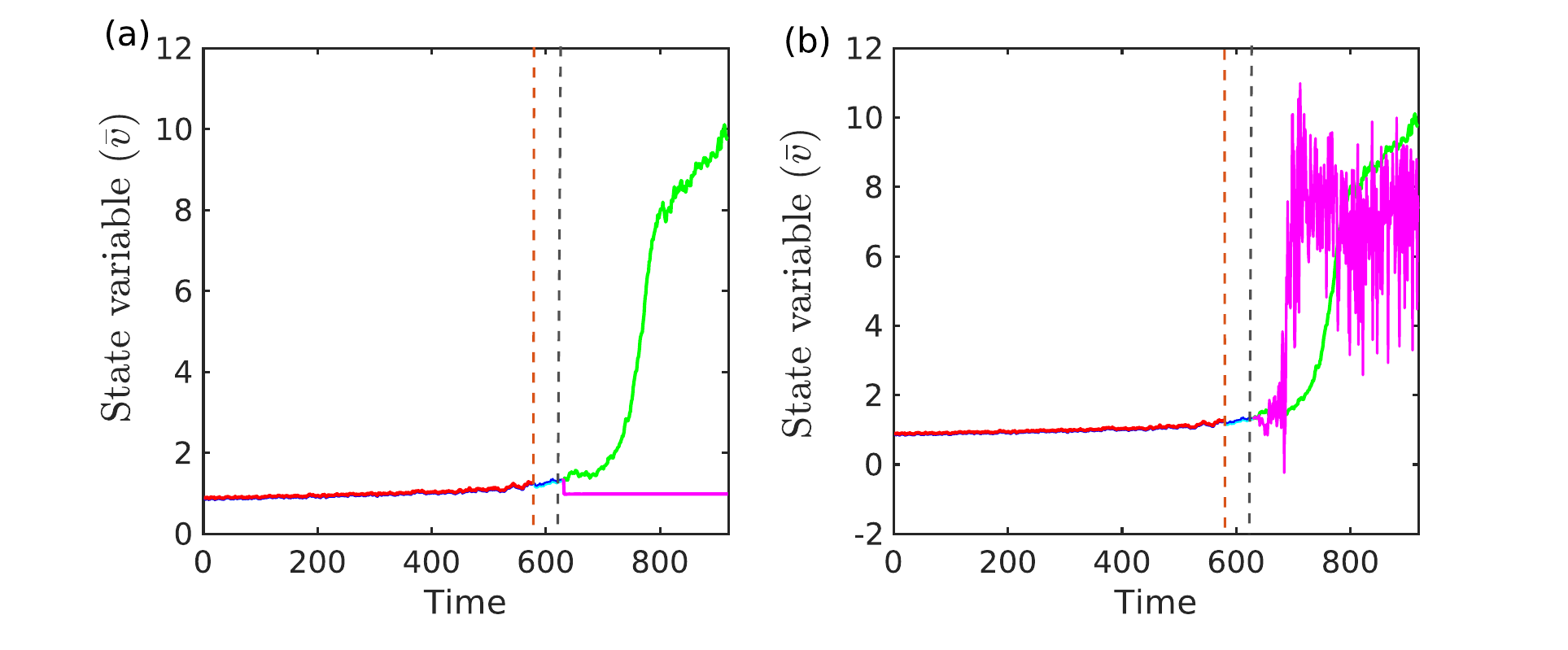}
\caption{Anticipating tipping in the vegetation-turbidity model, Eq.~\eqref{eq:VT}, using a time-delay feedforward network (TDFN). Panels show two representative outcomes: (a) unsuccessful anticipation, where no clear increasing trend is detected, and (b) successful anticipation, characterized by a pronounced increasing trend prior to the tipping point. The mean trajectories for the training, validation, and testing datasets are shown for the spatial system after applying NMF to the snapshots. The dashed orange and black vertical lines indicate the end of the training and the start of the testing periods, respectively. During the testing phase, the green curve denotes the ground truth.}
\label{fig:fig_fnn}
\end{figure}

To compare the performance of various methods in anticipating tipping, we evaluated alternative machine-learning methods~\cite{bishop2006pattern}, including time-delay feedforward networks (TDFNs) and convolutional neural networks (CNNs). We also tested classical spatial indicators of tipping points, including spatial standard deviation, spatial skewness, and spatial lag-1 correlation, using noisy data with varying noise amplitudes.

A TDFN is a standard feedforward neural network where inputs are augmented with delayed values of a time series. We constructed the TDFN with two hidden layers containing $100$ and $50$ hidden states, respectively; the delay is incorporated to better learn temporal dependencies. Similarly, the CNN is a more specialized type of feedforward network that learns patterns in data using convolution operations with learnable filters. We used a CNN with two convolution layers followed by a fully connected layer to obtain the final output at a given test time $t$. For both alternative machine-learning methods, the dimension of the spatial data was first reduced using NMF. The reduced data was then passed as input using a sliding window to serve as direct features for the TDFN and CNN. Figures~\ref{fig:fig_fnn} and \ref{fig:fig_cnn} show representative results of successful and unsuccessful anticipation by the TDFN and CNN methods, respectively. It can be seen that, while these models are unable to accurately estimate the exact tipping point, they are nevertheless capable of qualitatively anticipating an impending shift.

Traditionally, the phenomenon of tipping, characterized by a slowed recovery of a system to its previous state leading to a sudden shift in the system's state, can be manifested by changes in spatial statistics such as spatial standard deviation, spatial skewness, and lag-1 spatial correlation~\cite{kefi2014early}. More specifically, these measures are defined as follows.

Let $\bar{v}$ be the spatial mean of the state variable: $\bar{V}=\sum_{1}^{n_x} \sum_{1}^{n_y}v[x,y]/(n_x n_y)$, where $n_x$ and $n_y$ are the lattice dimensions along the $x$ and $y$ directions, respectively. Define the quantity $w[i,j;m,n]$, which takes a value of $1$ if positions $[i,j]$ and $[x,y]$ are at a distance $r$, and assumes a value of $0$ otherwise. Let $W$ be the total number of units separated by distance $r$. As the bifurcation point is approached, fluctuations in the state of the system increase, leading to enhanced variance:
\begin{align} \label{s1}
\sigma^2=\frac{1}{n_x n_y}\sum_{i=1}^{n_x}\sum_{j=1}^{n_y}(v[i,j]-\bar{v})^2,
\end{align}
The spatial standard deviation can then serve as an indicator of an approaching critical transition. Upon approaching a tipping point, the distribution of fluctuations about the mean spatial density becomes more asymmetric. This leads to observable changes in the third central moment modulated by the standard deviation, known as spatial skewness:
\begin{align} \label{s2}
\gamma=\frac{1}{n_x n_y} \sum_{i=1}^{n_x}\sum_{j=1}^{n_y} \frac{(v[i,j]-\bar{v})^3}{\sigma^3}.
\end{align}
Furthermore, in a spatial system, as a tipping point is approached, the spatial correlation---a measure of the linear dependence among neighboring units---also changes. Individual units tend to behave similarly, thereby increasing the spatial correlation prior to tipping. The spatial correlation is given by:
\begin{align} \label{s3}
C_2(r)=\frac{n_x n_y \sum_{i=1}^{n_x}\sum_{m=1}^{n_x}\sum_{j=1}^{n_y}\sum_{n=1}^{n_y} w[i,j;m,n](v[i,j]-\bar{v})(v[m,n]-\bar{v})}{W\sum_{m=1}^{n_x}\sum_{n=1}^{n_y}(v[m,n]-\bar{v})^2}.
\end{align}
Figure~\ref{fig:fig_ews} shows the spatial statistics obtained from the data corresponding to Eq.~\eqref{eq:VT}, with Kendall's-$\tau$ values listed in Tab.~\ref{tab:tab_tau} for the same pre-tipping data that was used for the machine-learning models. The trends in the spatial indicators are considered strong if the value of Kendall’s-$\tau$ coefficient is distributed closer to $1$. A pronounced increasing trend and higher Kendall’s-$\tau$ values are observed for data closer to the tipping point. Figure~\ref{fig:figbar} compares the results from the machine-learning models and those from the spatial statistics for the vegetation-turbidity system [Eq.~\eqref{eq:VT}] subject to noise of different amplitudes.

\begin{figure}[ht!]
\centering
\includegraphics[width=1.0\linewidth]{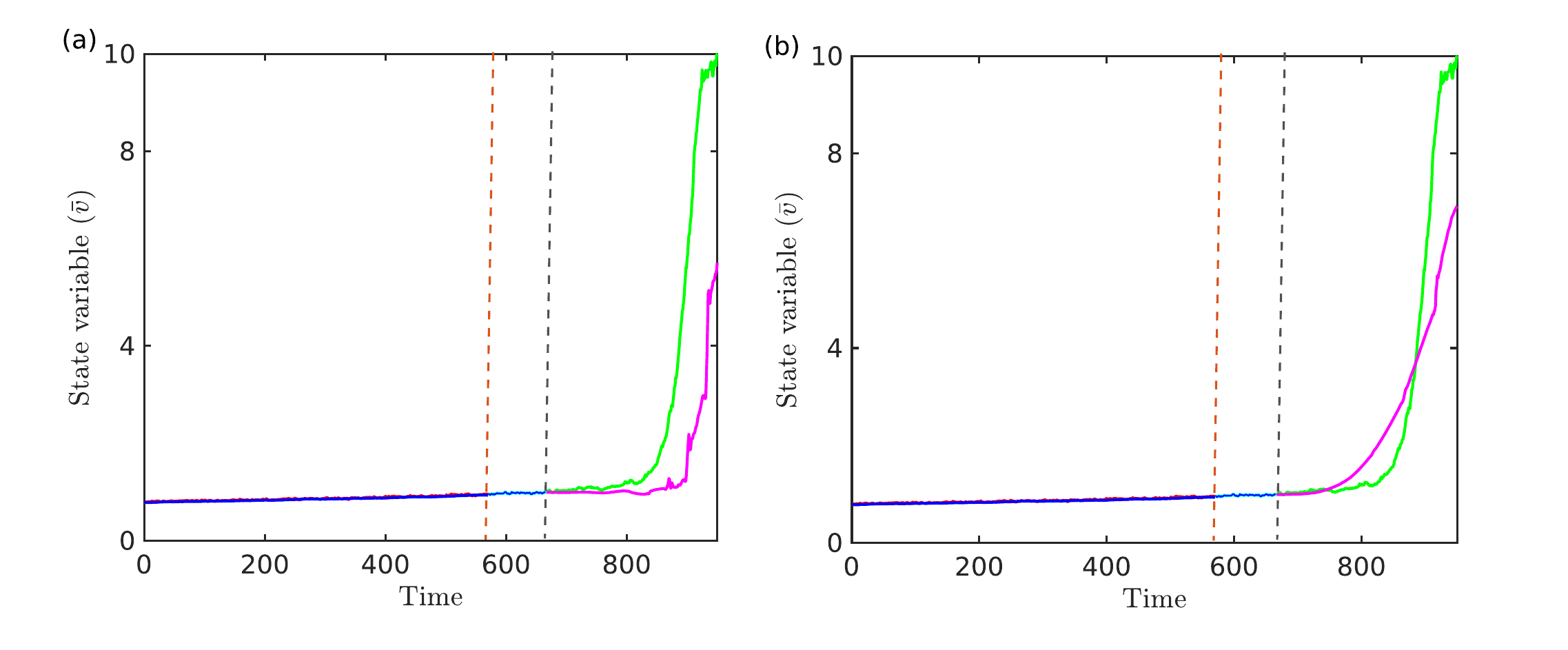}
\caption{Anticipating tipping in the vegetation-turbidity model, Eq.~\eqref{eq:VT}, with a convolutional neural network (CNN). Two representative outcomes: (a) unsuccessful anticipation, where no clear increasing trend is detected prior to tipping, and (b) successful anticipation, characterized by a pronounced increasing trend prior to the tipping point. The mean trajectories for the training, validation, and testing datasets are shown for the spatial system after applying NMF to the snapshots. The dashed orange and black vertical lines indicate the end of the training and the start of the testing periods, respectively. During the testing phase, the green curve denotes the ground truth.}
\label{fig:fig_cnn}
\end{figure}

\begin{figure}[ht!]
\centering
\includegraphics[width=1\linewidth]{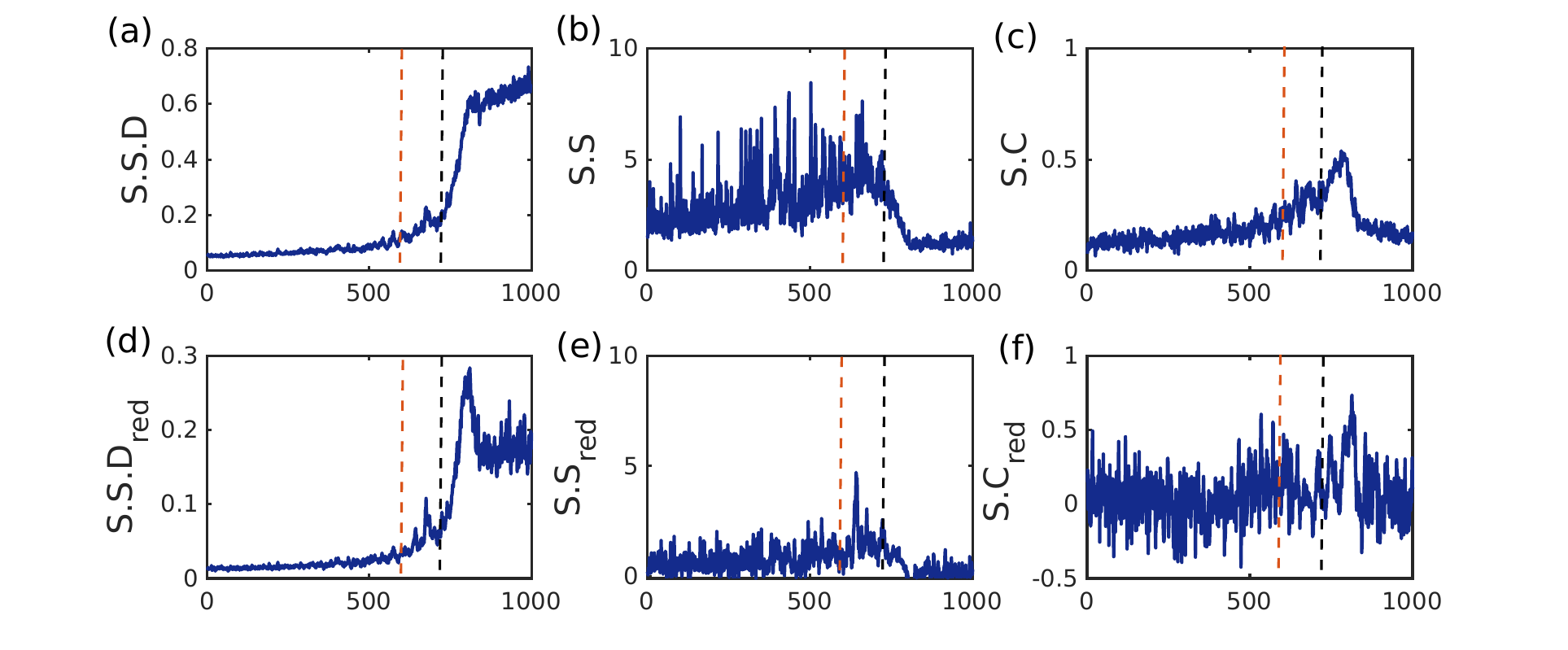}
\caption{Trends in spatial indicators for the vegetation-turbidity model, Eq.~\eqref{eq:VT}. (a–c) Spatial standard deviation, spatial skewness, and spatial lag-1 correlation, respectively, computed from the original system. (d–f) The corresponding indicators for the reduced system obtained via coarse-graining using $5\times 5$ submatrices. The orange and black dashed vertical lines indicate the time interval over which the spatial indicators are used to compute Kendall's-$\tau$ to compare the strength of the observed trends.}
\label{fig:fig_ews}
\end{figure}


\begin{table}[ht!]
\caption{Kendall's-$\tau$ values corresponding to the orange ($\tau_{1}$) and black ($\tau_{2}$) dashed lines, respectively, in Fig.~\ref{fig:fig_ews}. The values are presented for the three spatial indicators computed from snapshots along the gradients of the control parameter by solving Eq.~\eqref{eq:VT}.}
\label{tab:tab_tau}
\begin{tabular}{|c|c|c|}
		\hline
		Spatial indicator &  $\tau_1$ &  $\tau_2$ \\
		\hline
		S.S.D   &   0.79        &    0.85       \\
		S.S  &        0.43   &     0.34     \\
		S.C    &      0.54     &       0.68    \\
		$S.S.D_{red}$    & 0.72          &   0.80        \\
		$S.S_{red}$   &      0.27     &    0.29       \\
		$S.C_{red}$  &     0.12      &      0.17     \\
		\hline
\end{tabular}
\end{table}

\begin{table}[ht!]
\caption{Wilson confidence interval ($CI_{W}$) computed for predictions from $1000$ realizations of the parameter-adaptable reservoir computer, using spatial data simulated at different noise amplitudes. This interval provides an estimate of the fraction of realizations in which the predicted tipping point falls within the true tipping interval, calculated at a $95\%$ confidence level.} 
\label{tab:tab_RC_noise}
\begin{tabular}{|c|c|c|}
		\hline
		Noise amplitude ($\sigma$) &  $CI_{W}$\\
		\hline
		0.05  &   [83.8,88]       \\
		0.1  &    [82.5,87]    \\
		0.15    &  [81,86]      \\
		0.2    &  [77.4,82]  \\
		0.25  &   [73.2,78.5] \\
		0.3 &     [69.5,75]  \\
		\hline
\end{tabular}
\end{table}


\section{Threshold generalized additive model} \label{secS2}

It is worth comparing parameter-adaptable reservoir computing with threshold generalized additive models (TGAMs)~\cite{andersen2009ecological,bestelmeyer2011analysis} for detecting breakpoints in time series. When applied to time series data containing a critical threshold or tipping point, TGAMs can, in principle, determine the exact location of the transition. The implementation of a TGAM involves the following steps.

Consider a time series dataset
\begin{equation*}
\{(t_i, y_i)\}_{i=1}^{N},
\end{equation*}
where the system undergoes a qualitative change in behavior at an unknown time $\theta$, identified as the tipping point associated with bistability and hysteresis. The objective is to estimate $\theta$ such that the time series is optimally separated into two distinct regimes. Assume that the observed signal can be represented as a piecewise smooth function,
\begin{equation*}
y_i =
\begin{cases}
f_1(t_i) + \varepsilon_i, & t_i \le \theta, \\
f_2(t_i) + \varepsilon_i, & t_i > \theta,
\end{cases}
\end{equation*}
where $f_1$ and $f_2$ are smooth nonlinear functions, and $\varepsilon_i$ denotes the residual error. Each smooth function $f_j(t)$ ($j \in \{1,2\}$) can be estimated by minimizing the penalized least-square functional
\begin{equation*}
f_j = \arg\min_{g}
\left[
\sum_{t_i \in R_j} \left(y_i - g(t_i)\right)^2
+ \beta_j \int \left(g''(t)\right)^2 \, dt
\right],
\end{equation*}
where $R_1 = \{t_i : t_i \le \theta\}$ and $R_2 = \{t_i : t_i > \theta\}$ denote the two regimes, and $\beta_j$ controls the smoothness penalty. Specifically, given a set of candidate thresholds $\{\theta_k\}$, the optimal threshold can be determined as follows:
\begin{enumerate}
\item
        Split the data into two regions:
        $$
        (t_i, y_i) =
        \begin{cases}
        \text{Region 1}, & t_i \le \theta_k, \\
        \text{Region 2}, & t_i > \theta_k.
        \end{cases}
        $$
\item
        Fit independent smoothing functions $f_1$ and $f_2$ on the two regions.
\item
        Reconstruct the signal:
        \begin{equation*}
        \hat{y}_i(\theta_k) =
        \begin{cases}
        f_1(t_i), & t_i \le \theta_k, \\
        f_2(t_i), & t_i > \theta_k.
        \end{cases}
        \end{equation*}
\item
        Compute the residual sum of squares:
        \begin{equation*}
        \mathrm{RSS}(\theta_k) =
        \sum_{i=1}^{N} \left(y_i - \hat{y}_i(\theta_k)\right)^2.
        \end{equation*}
\item
        Next, evaluate the Akaike Information Criterion:
        \begin{equation*}
        \mathrm{AIC}(\theta_k) =
        N \ln\!\left(\frac{\mathrm{RSS}(\theta_k)}{N}\right) + 2k,
        \end{equation*}
where $k$ ($=1$, for determining a tipping point) is the number of estimated threshold parameters.
\end{enumerate}
The optimal tipping point is given by
\begin{equation*}
\theta^\ast = \arg\min_{\theta_k} \mathrm{AIC}(\theta_k).
\end{equation*}
We implement the TGAM to estimate tipping points from both the original and reservoir-computer-predicted time series. The results from the NMF-reduced, spatially averaged system closely match those derived from the true time series. Representative results are shown in Figs.~\ref{fig:figS3}-\ref{fig:figS8}.

\begin{figure} [ht!]
\centering
\includegraphics[width=0.45\linewidth]{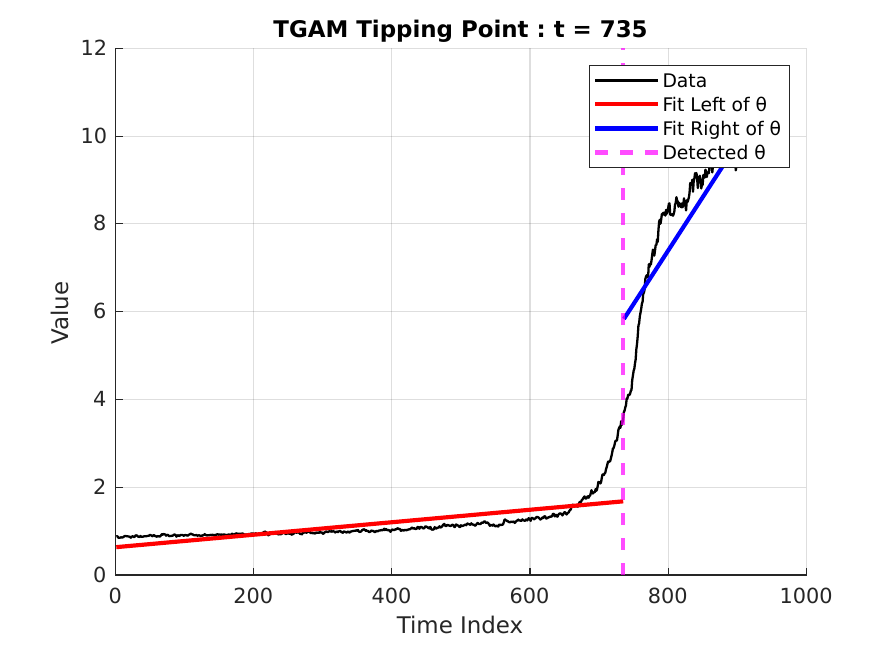}
\includegraphics[width=0.384\linewidth]{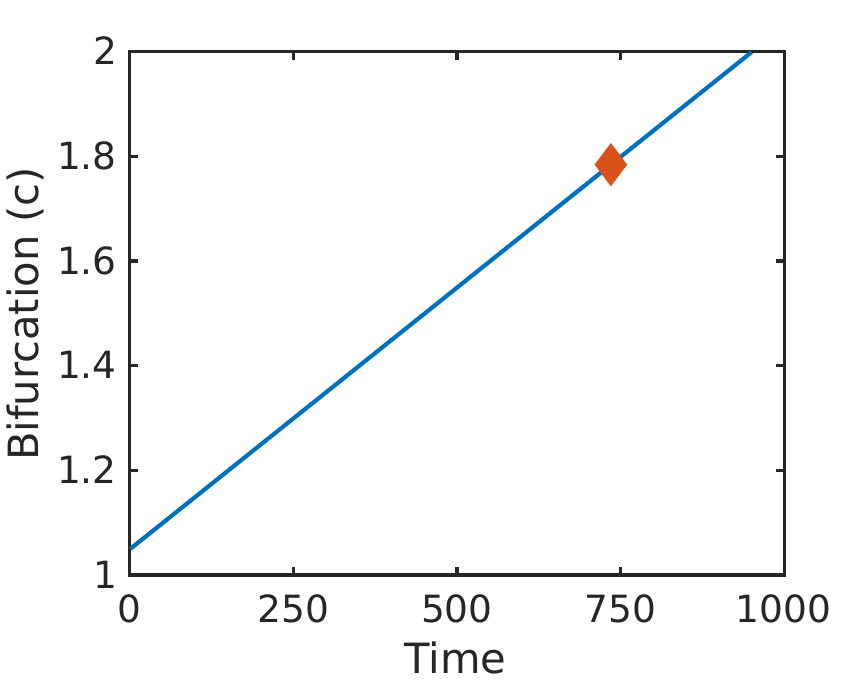}
\caption{Estimating the tipping point of the vegetation-turbidity system using a TGAM. (a) Estimated tipping point for the vegetation-turbidity model, Eq.~\eqref{eq:VT}. (b) The bifurcation parameter as a function of time. The orange marker indicates the true tipping threshold of the bifurcation parameter ($c=1.784$ at $t=735$) as estimated using the TGAM.}
\label{fig:figS3}
\end{figure}

\begin{figure} [ht!]
\centering
\includegraphics[width=0.45\linewidth]{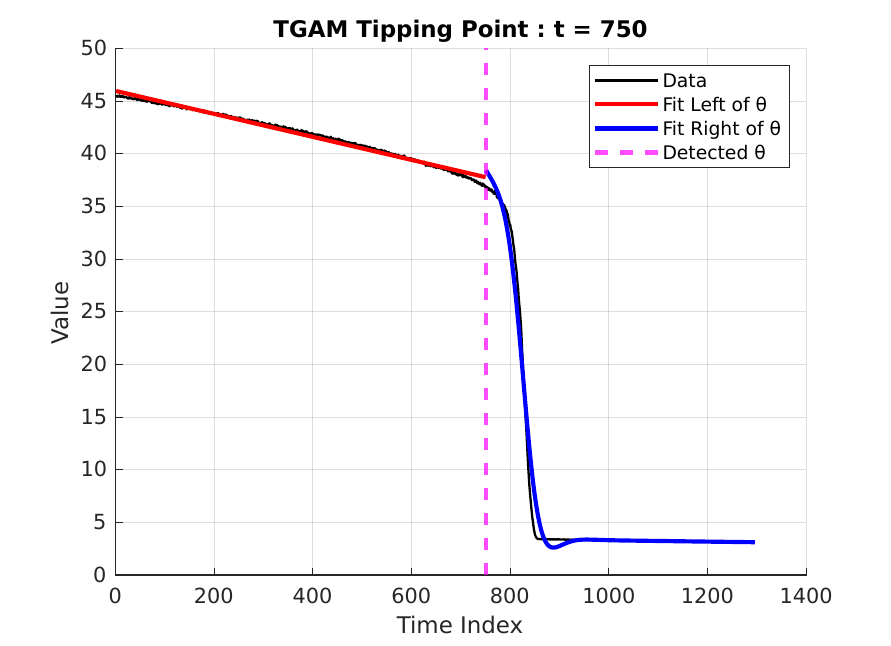}
\includegraphics[width=0.384\linewidth]{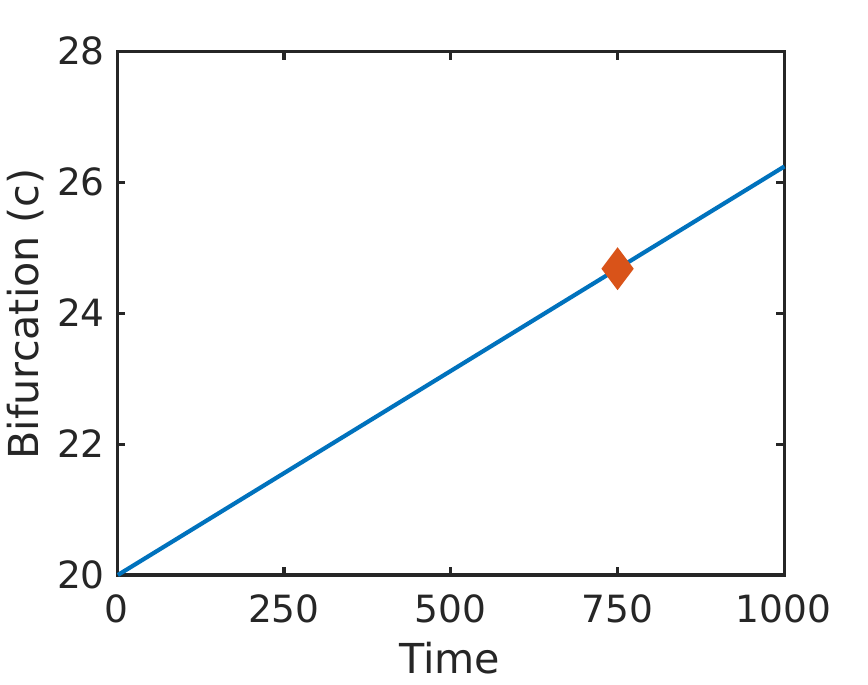}
\caption{Estimating the tipping point of the vegetation-grazing system using a TGAM. (a) Estimated tipping point for the vegetation-grazing model, Eq.~\eqref{eq:VG}. (b) The bifurcation parameter as a function of time. The orange marker indicates the true tipping threshold of the bifurcation parameter ($c=24.6812$ at $t=750$) as estimated using the TGAM.}
\label{fig:figS4}
\end{figure}

\begin{figure} [ht!]
\centering
\includegraphics[width=0.45\linewidth]{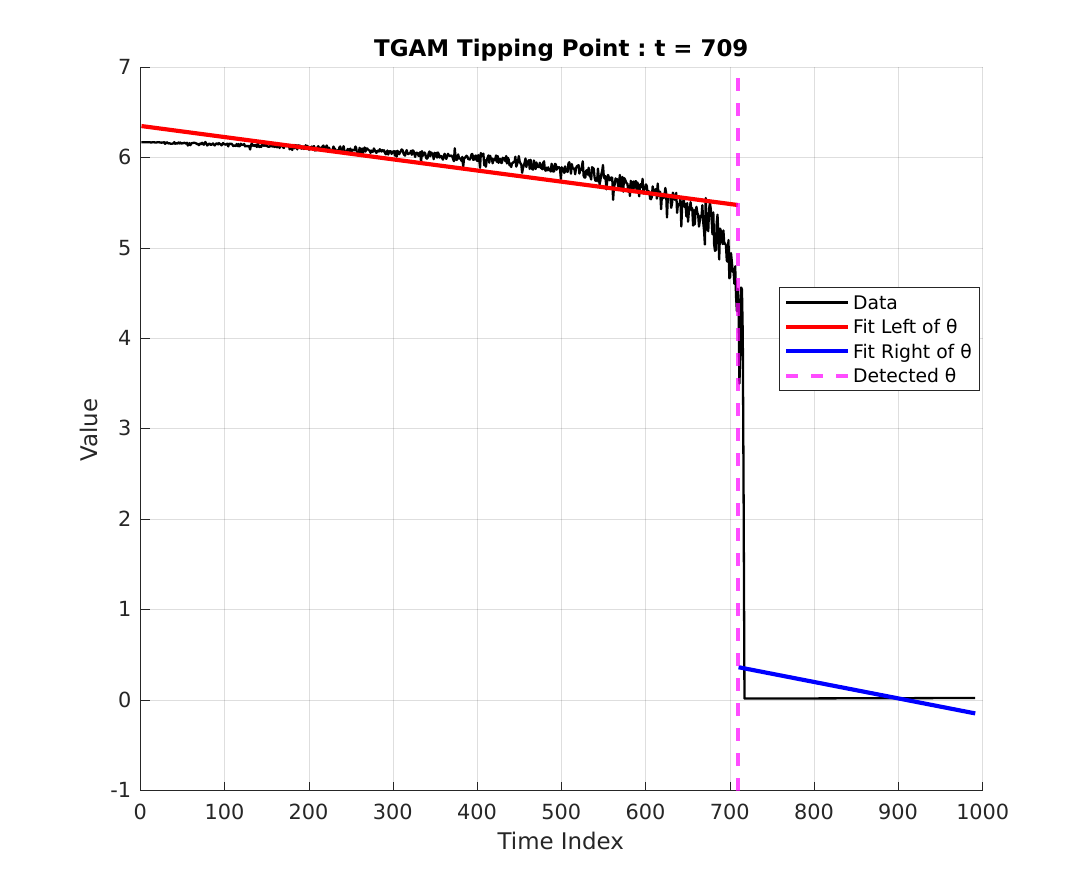}
\includegraphics[width=0.384\linewidth]{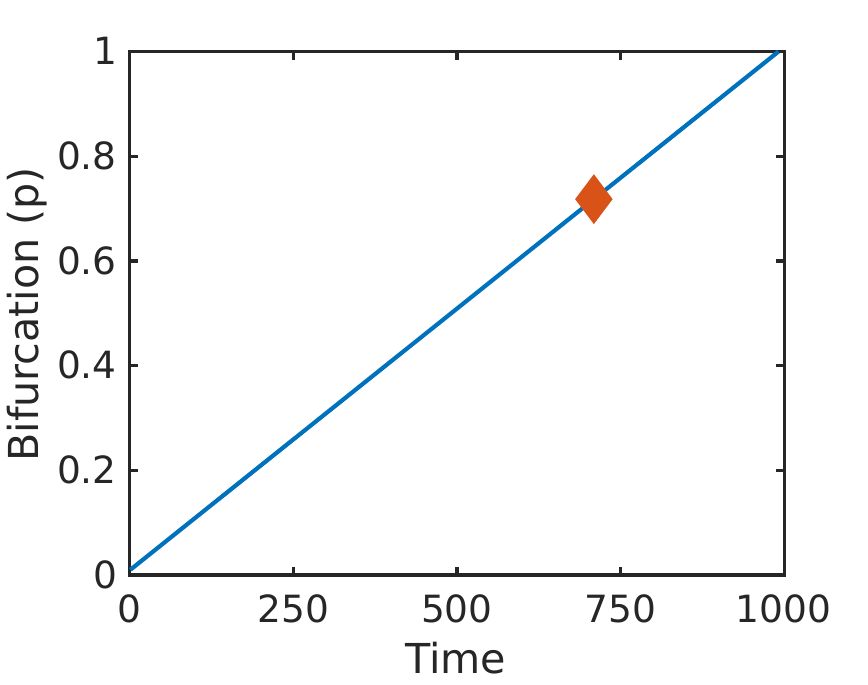}
\caption{Estimating the tipping point of the cellular automata model using a TGAM. (a) Estimated tipping point for the cellular automata model, Eq.~\eqref{eqCA}. (b) The bifurcation parameter as a function of time. The orange marker indicates the true tipping threshold of the bifurcation parameter ($c=0.718$ at $t=709$) as estimated using the TGAM.}
\label{fig:figS5}
\end{figure}

\begin{figure} [ht!]
\centering
\includegraphics[width=0.5\linewidth]{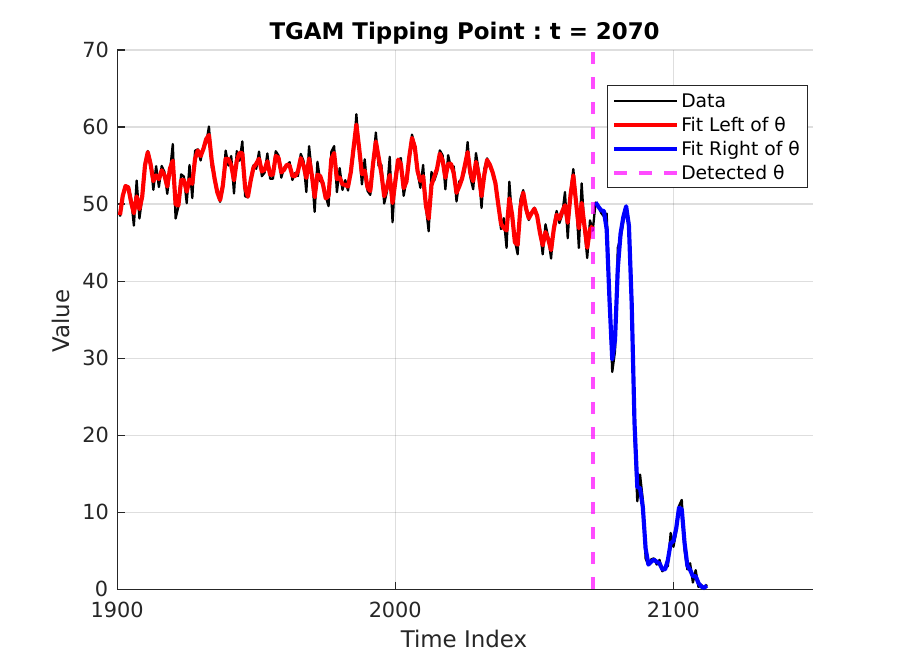}
\caption{Estimating the tipping point of the first CMIP5 dataset. Shown is the tipping point estimated using a TGAM for the CMIP5 projected climate variable: percentage sea ice cover in the high-latitude oceans from MRI‑CGCM3.}
\label{fig:figS6}
\end{figure}

\begin{figure}[ht!]
\centering
\includegraphics[width=0.5\linewidth]{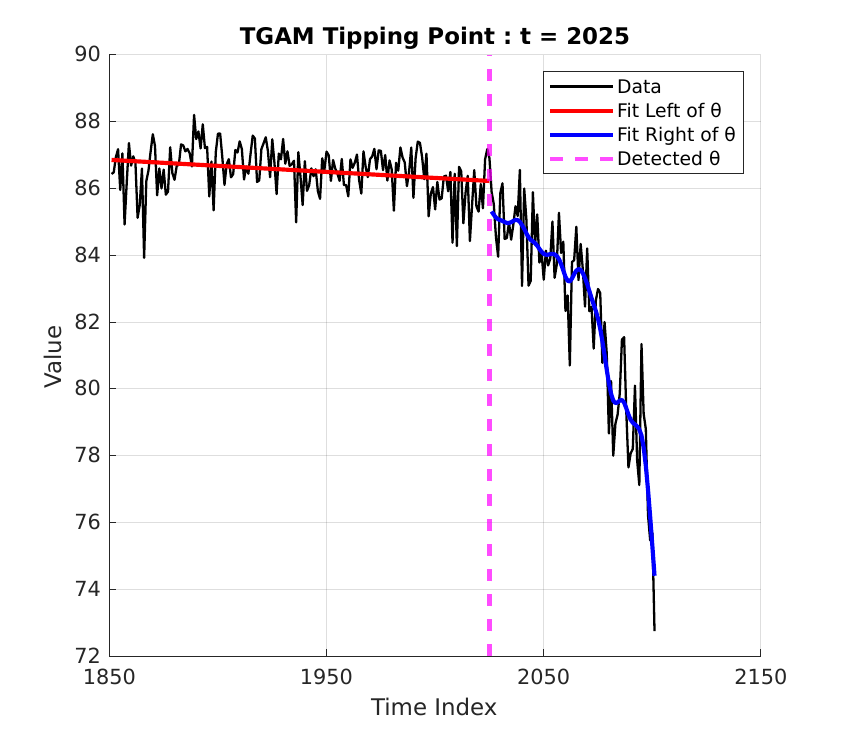}
\caption{Estimating the tipping point of the second CMIP5 dataset. Shown is the tipping point estimated using a TGAM for the CMIP5 projected climate variable: percentage sea ice cover in the Arctic Ocean from CSIRO-MK3-6-0.}
\label{fig:figS7}
\end{figure}

\begin{figure}[ht!]
\centering
\includegraphics[width=0.5\linewidth]{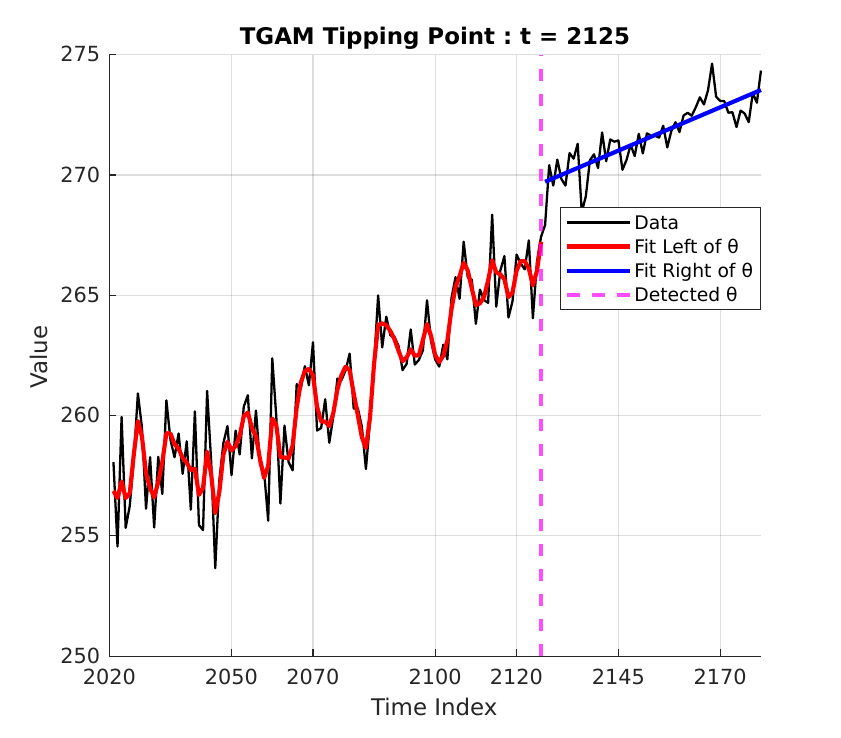}
\caption{Estimating the tipping point of the third CMIP5 dataset. Shown is the tipping point estimated using a TGAM for the CMIP5 projected climate variable: temperature in the month of April for MPI-ESM-LR for the Pacific sector of the Arctic Ocean.}
\label{fig:figS8}
\end{figure}

\section{Reservoir-computing prediction for individual spatial cells} \label{secS3}

Figures~\ref{fig:figS9}-\ref{fig:figS14} and Tabs.~\ref{tab:tab1} and \ref{tab:tab2} present the results of the reservoir-computing predictions at the level of individual spatial cells. Our consistent observations across all spatial cells indicate that the reservoir computing framework uniformly predicts abnormal behavior across all spatial cells in diverse systems. This ensures that the averaged prediction plots in the main draft (Fig. \ref{fig:VT_model}, \ref{fig:Grazing}, \ref{fig:CMIP5}) are not artifacts or occur due to chance arising from aggregation or cancellation effects at the level of individual cells.

\begin{figure} [ht!]
\centering
\includegraphics[width=\linewidth]{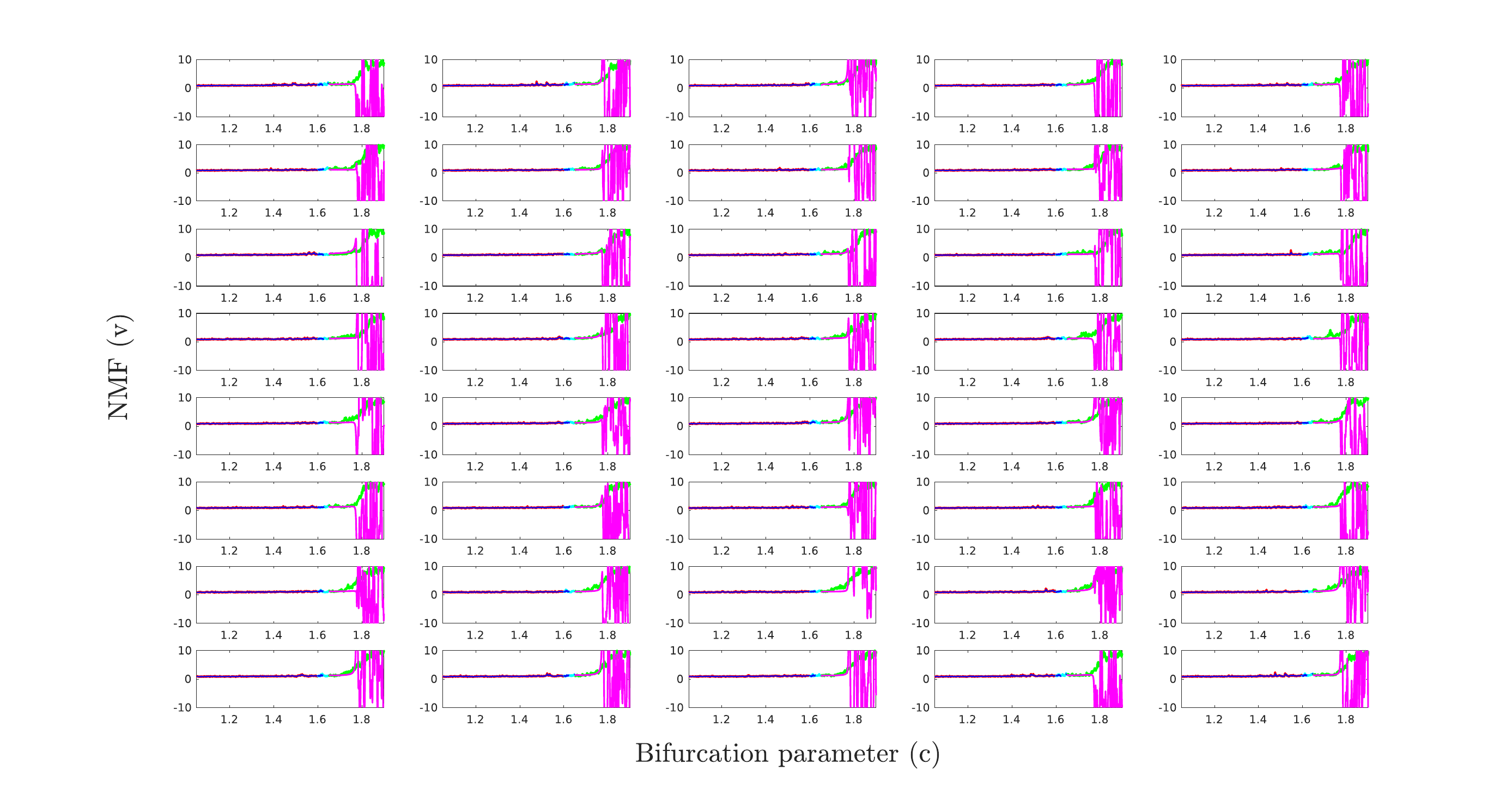}
\caption{Training, validation, and testing by the reservoir computer on the spatial vegetation-turbidity system following the application of NMF to the data snapshots for each spatial cell. The mean values are presented in Fig.~\ref{fig:VT_model}(b).}
\label{fig:figS9}
\end{figure}

\begin{figure} [ht!]
\centering
\includegraphics[width=\linewidth]{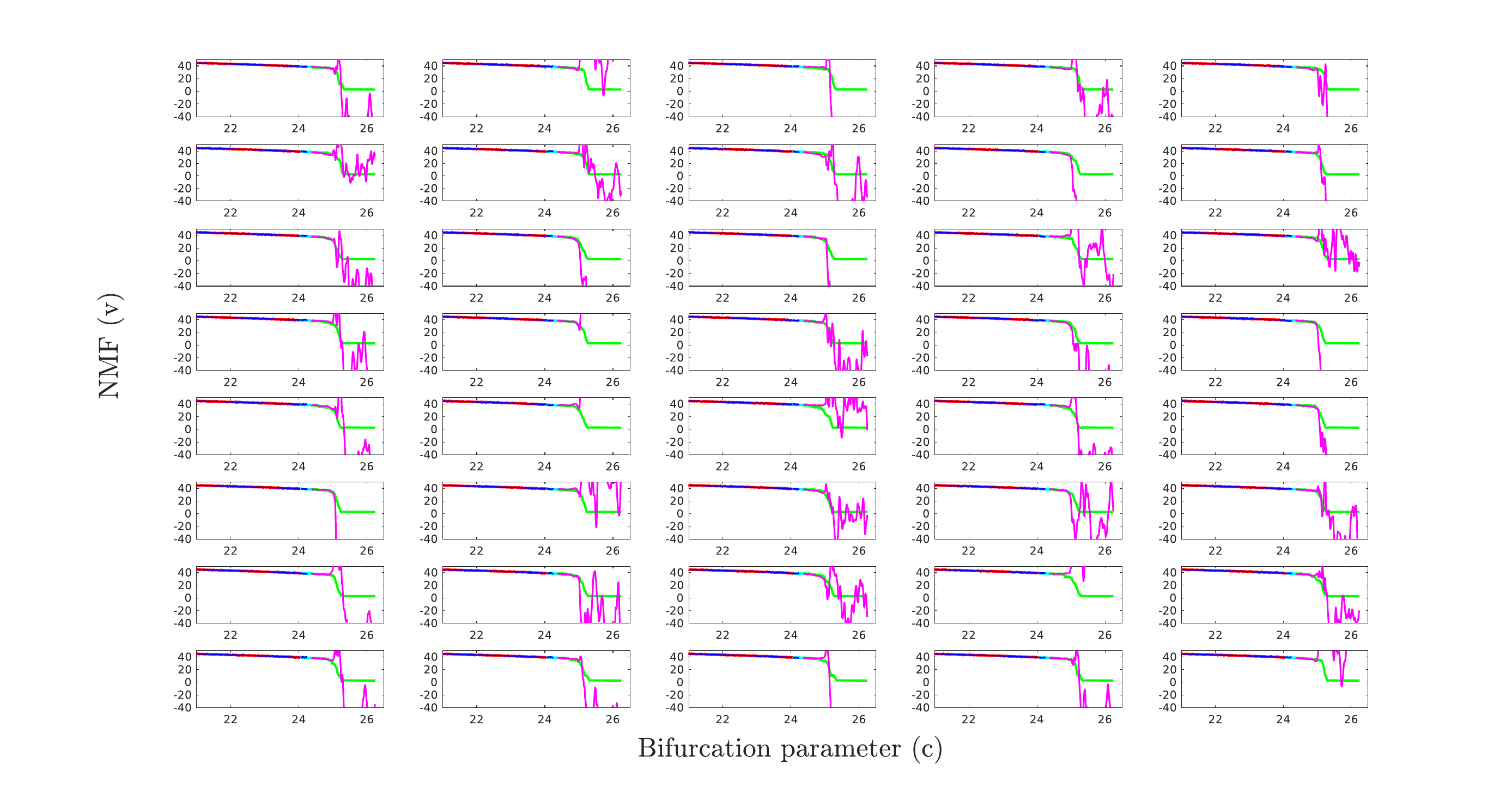}
\caption{Training, validation, and testing by the reservoir computer on the spatial vegetation-grazing system following the application of NMF to the data snapshots for each spatial cell. The mean values are shown in Fig.~\ref{fig:Grazing}(b).}
\label{fig:figS10}
\end{figure}

\begin{figure} [ht!]
\centering
\includegraphics[width=\linewidth]{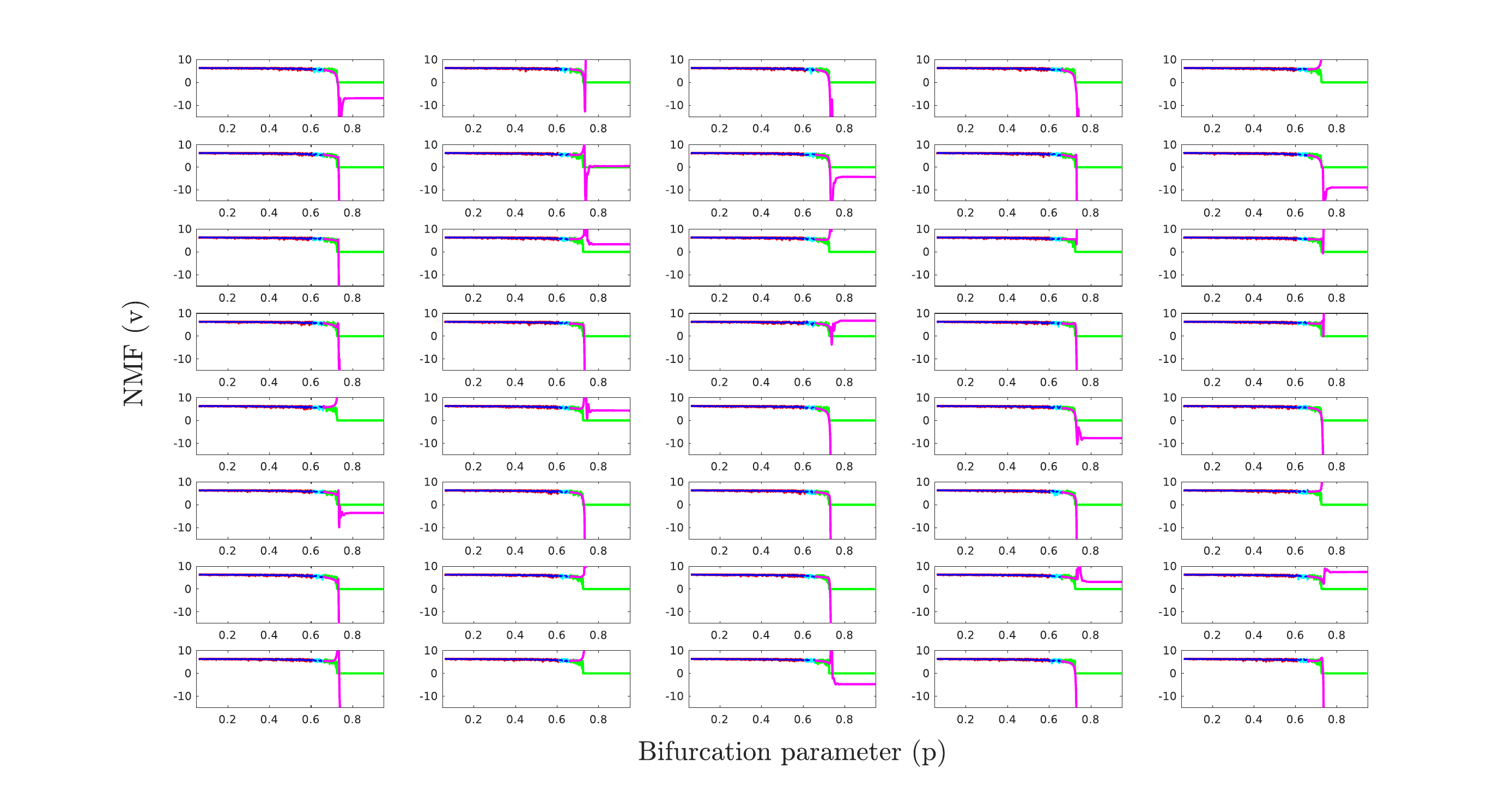}
\caption{Training, validation, and testing by the reservoir computer on the spatial cellular automata system following the application of NMF to the data snapshots for each spatial cell. The mean values are presented in Fig.~\ref{fig:Grazing}(f).}
\label{fig:figS11}
\end{figure}

\begin{figure} [ht!]
\centering
\includegraphics[width=\linewidth]{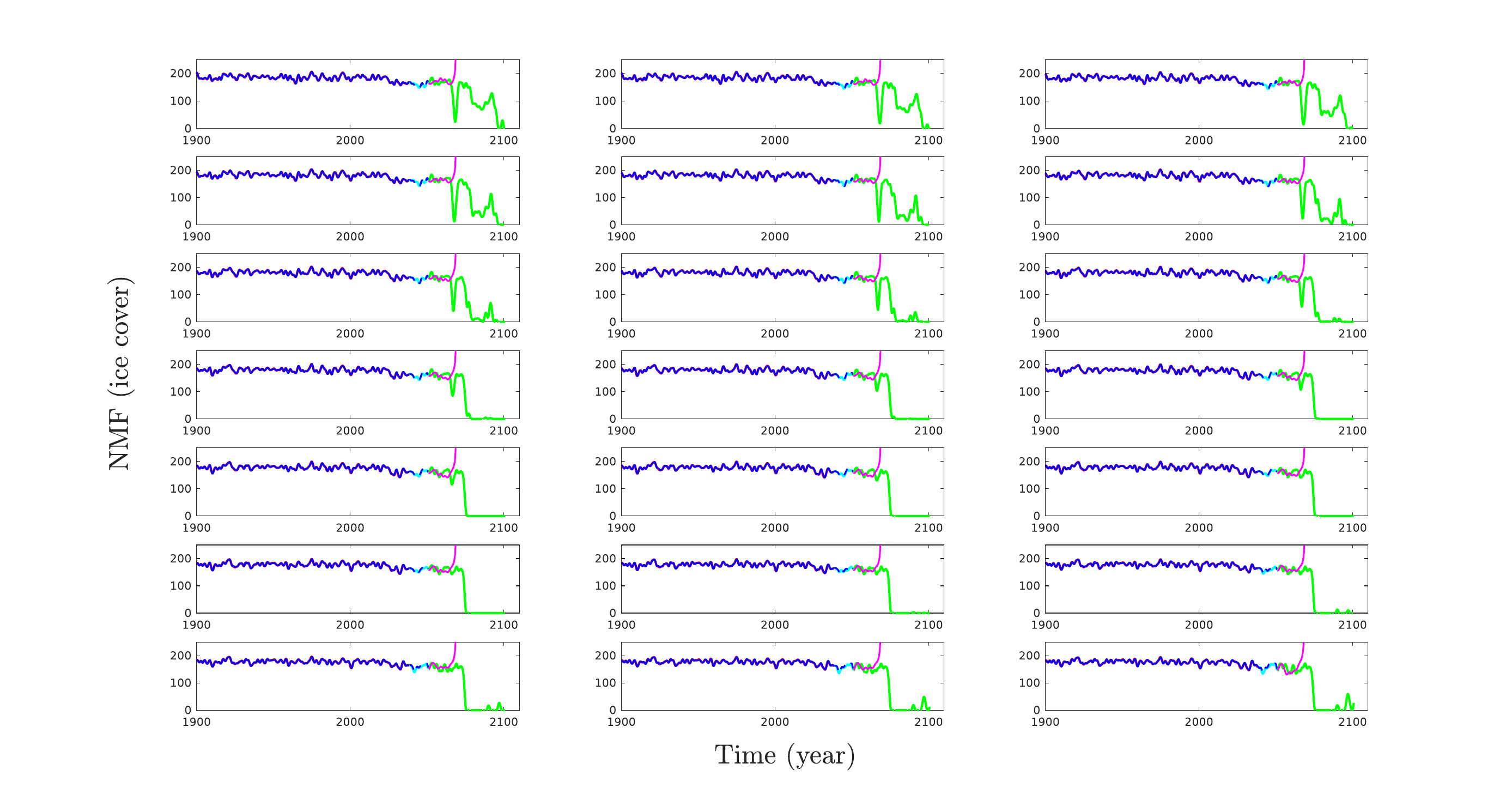}
\caption{The NMF-reduced sea ice cover in the high latitudes from MRI‑CGCM3 versus time for each spatial cell. Plots are displayed for the training, validation, and testing datasets of the spatial system following the application of NMF to the data snapshots. The mean values for the training, validation, and testing datasets of the spatial system after applying NMF to the snapshots are shown in Fig.~\ref{fig:CMIP5}(b).}
\label{fig:figS12}
\end{figure}

\begin{figure} [ht!]
\centering
\includegraphics[width=\linewidth]{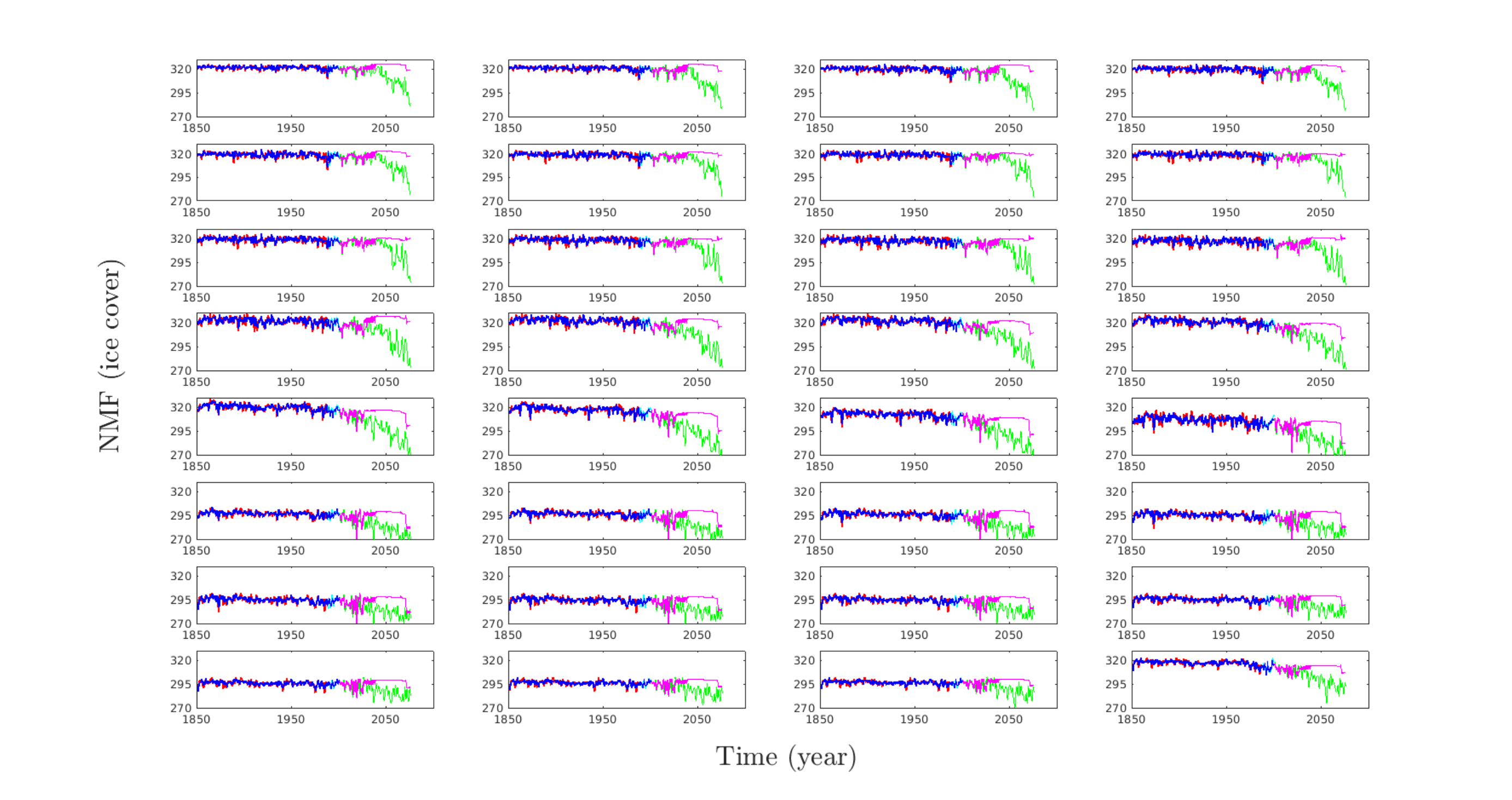}
\caption{The NMF-reduced annual sea ice cover in the Arctic Ocean from CSIRO-MK3-6-0 versus time for each spatial cell. Plots are displayed for the training, validation, and testing datasets of the spatial system following the application of NMF to the data snapshots. The mean values for the training, validation, and testing datasets of the spatial system after applying NMF to the snapshots are presented in Fig.~\ref{fig:CMIP5}(f).}
\label{fig:figS13}
\end{figure}

\begin{figure} [ht!]
\centering
\includegraphics[width=\linewidth]{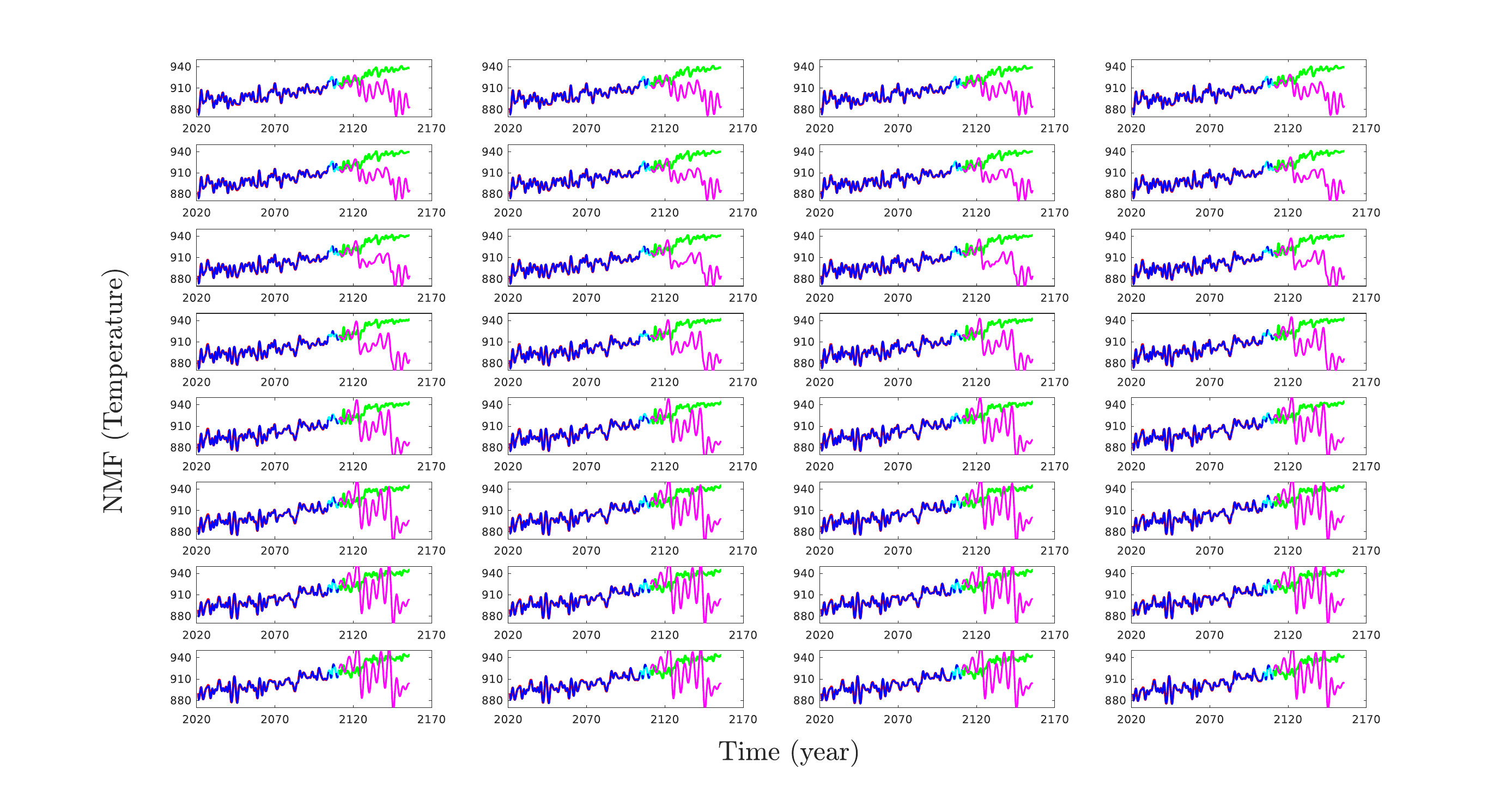}
\caption{The NMF-reduced temperature in the month of April for MPI-ESM-LR for the Pacific sector of the Arctic Ocean versus time for each spatial cell. Plots are displayed for the training, validation, and testing datasets of the spatial system following the application of NMF to the data snapshots. The mean values for the training, validation, and testing datasets of the spatial system after applying NMF to the snapshots are presented in Fig.~\ref{fig:CMIP5}(j).}
\label{fig:figS14}
\end{figure}

\begin{table} [htbp]
\centering
\caption{Hyperparameter values for the spatial data across the different models used in the reservoir-computing predictions. The size of the reservoir network is $2000$ for all cases.}
\label{tab:tab1}
\begin{tabular}{|c|c|c|c|c|c|c|}
\hline
Spatial Data & $\rho$ &$\gamma$  & $\alpha$ & $\beta$ & $d$  & $b_{0}$ \\
\hline
Vegetation turbidity model & 0.682 & 0.0235  & 0.572  &  $10^{-3.9404}$& 0.2535 & 1.4794 \\
\hline
Vegetation grazing model & 0.637 & 0.0235 &  0.1523&$10^{-3.3672} $ & 0.247 & 0.5252 \\
\hline
Cellular automata model & 0.4713 &  0.02829 & 0.4147  & $10^{-3.1260}$ & 0.2164 & 0.7245 \\
\hline
Temperature (MPI-ESM-LR)& 0.68  &0.0007  & 0.57 & $10^{-3.5}$ & 0.2535 &  0.1\\
\hline
$\%$ sea ice cover (CSIRO-MK3-6-0) & 0.6 & 4.009  & 0.65 & $10^{-3.4.45}$ & 0.25 & 1 \\
\hline
$\%$ sea ice cover (MRI-CGCM3)& 0.65 &0.0029  &0.5965  & $10^{-3.65}$ &0.35  &3  \\
\hline
No transition &0.4905  & 3.0056 &  0.8328& $10^{-3.1632} $ &0.5174  &0.9536  \\
\hline
\end{tabular}
\end{table}

\begin{table} [htbp]
\centering
\caption{Details of the CMIP5 climate projection data sampling used to evaluate the generality of the parameter-adaptable reservoir computer on real-world datasets. All datasets are sampled at a monthly resolution. For the MPI-ESM-LR data, only the temperature data for the month of April is used. For all remaining cases, monthly data for all $12$ months are sampled and averaged to obtain the annual mean values of the respective variables. Data from $1850–2005$ correspond to the historical experiment, while data from $2006–2100$ correspond to the RCP scenario specified for each case in the table.}
\label{tab:tab2}
\begin{tabular}{|c|c|c|c|}
\hline
Details & CMIP5 data 1  & CMIP5 data 2  & CMIP5 data 3  \\
\hline
Model & MPI-ESM-LR & CSIRO-MK3-6-0 & MRI-CGCM3 \\
\hline
Variable & ta & sic &  sic\\
\hline
Scenario & RCP85 & RCP85  & RCP26 \\
\hline
Time & 2020-2100  &1850-2100  & 1850-2100 \\
\hline
Time frequency & mon &mon  & mon \\
\hline
Ensemble & r1i1p1 & r5i1p1 & r1i1p1 \\
\hline
Region & Pacific region of Arctic Ocean & Arctic Ocean & High latitude ocean \\
\hline
\end{tabular}
\end{table}

\end{widetext}

\end{document}